\documentclass[12pt]{article}
\usepackage[dvips]{epsfig}
\usepackage{latexsym}
\usepackage{caption}  % Customize captions
\newcommand{\be}{\begin{equation}}
\newcommand{\ee}{\end{equation}}

\def\bbox{{\,\lower0.9pt\vbox{\hrule \hbox{\vrule height 0.2 cm
\hskip 0.2 cm
\vrule  height 0.2 cm}\hrule}\,}}

\newcommand{\beq}{\begin{equation}}
\newcommand{\eeq}{\end{equation}}
\newcommand{\bea}{\begin{eqnarray}}
\newcommand{\eea}{\end{eqnarray}}

\begin{document}
\setlength{\unitlength}{1mm}
\title{{\hfill {\small } } \vspace*{2cm} \\
Thorny Spheres and Black Holes with Strings}
\author{\\
V.P. Frolov${}^1$, D.V. Fursaev${}^2$, D.N. Page${}^1$}
\maketitle
\noindent  {
$^{1}${ \em
Theoretical Physics Institute, Department of Physics, \ University of
Alberta, \\ Edmonton, Canada T6G 2J1;
\\ Fellow, Canadian Institute for Advanced Research}
\\ $^{2}${\em Joint Institute for
Nuclear Research,
Bogoliubov Laboratory of Theoretical Physics,\\
 141 980 Dubna, Russia}
\\
e-mails: frolov@phys.ualberta.ca, fursaev@thsun1.jinr.ru,
don@phys.ualberta.ca
}
\bigskip

\maketitle %%%%%%%%%%%%%%%%%%%%%%%%%%%%%%%%%%%%%%%%%%%%%%%%%%%%%%%
\begin{abstract} 
We consider thorny spheres, that is  2-dimensional compact surfaces
which are everywhere locally isometric to a round sphere $S^2$ except
for a finite number of isolated points where they have conical
singularities. We use thorny spheres to generate, from a spherically
symmetric solution of the Einstein equations,  new solutions which
describe  spacetimes  pierced by an arbitrary number of infinitely thin
cosmic strings radially directed.  Each string produces an angle
deficit proportional to its tension,  while the metric outside the
strings is a locally spherically symmetric solution.  We prove that
there can be arbitrary configurations of strings provided that the
directions of the strings obey a certain equilibrium condition. In
general this equilibrium condition can be written as a force-balance
equation for string forces defined in a flat 3-space in which the
thorny sphere is isometrically embedded, or as a constraint on the
product of holonomies around strings in an alternative 3-space that is
flat except for the strings. In the case of small string tensions, the
constraint equation has the form of a linear relation between unit
vectors directed along the string axes. 
\end{abstract}
%%%%%%%%%%%%%%%%%%%%%%%%%%%%%%%%%%%%%%%%%%%%%%%%%%

\bigskip
%\vspace{3cm}

\bigskip

\baselineskip=.6cm

\newpage

\section{Introduction}
\setcounter{equation}0

Recently \cite{FrFu:00} it was demonstrated that cosmic strings
attached radially to a black hole can be used for very effective energy
mining from black holes. There were also found sets of exact solutions
of the Einstein equations which describe a black hole with infinitely
thin radial  cosmic strings \cite{FrFu:01} and generalize results
of \cite{AFV},\cite{DoCh:92}. 
For such solutions a
regular round sphere is changed to a sphere with a number of conical
singularities on it with angle deficits $\mu = 8\pi\tilde{\mu}$,
where $\tilde{\mu}$ is the dimensionless\footnote
{We work in the system of units $\hbar=G=c=1$. In
these units the string tension $\tilde{\mu}$ is dimensionless
and corresponds to the combination $G\hat{\mu}/c^2$, where $\hat{\mu}$
is the tension measured in physical units.}
cosmic string tension.  

A characteristic property of the configurations studied in
\cite{FrFu:01} is that the positions of the conical singularities on a
sphere form a regular symmetric structure. The number of types of these
configurations is restricted. There are three configurations which are
related to Platonic solids and one family of configurations which looks
like a `double pyramid'. In the latter case a number of conical
singularities (and hence the strings) is not restricted.

In physical applications one can always assume that the string tension
is very small. For example, for strings which appear in GUT  theories
the tension is $10^{-6}$,
while for electroweak strings it is $10^{-34}$.
Finding all possible static radial string configurations for small $\mu$
without any additional a priori symmetry assumptions is our first goal
in the present paper.
We shall demonstrate that such configurations exist
for any number $n < 4\pi/\mu$ of strings,
and in the general case they do not possess any symmetry.
Nevertheless there always exists a vector force-balance
constraint equation
\begin{equation}\label{i0}
{\bf F} = \sum_{k=1}^n {\bf F}_k = 0\, ,
\end{equation}
which for $\sum_{i=1}^n \mu_k \ll 1$
is approximated by
\begin{equation}\label{i1}
\sum_{k=1}^n \mu_k {\bf n}_k=0\, .
\end{equation}
The sum is taken over all singular points with corresponding angular
deficits $\mu_k$. In the given approximation the position of each point
is characterized by the unit vector ${\bf n}_k$
on a smooth sphere $S^2$.

We studied also in detail the case when the string tension is not small,
and this is our second main goal. We demonstrated that there can exist
configurations with an arbitrary number of strings $n$, provided the
total angular deficit is less than $4\pi$. The constraint equations
which again must be satisfied are now more involved. We demonstrate
that these relations can be written as a constraint on the products of
the elements of the holonomy group representing the conical
singularities. The constraint equations are more involved since the
corresponding operators of the holonomy group do not commute. 

This analysis requires knowledge of different geometrical properties of
an object which we called a {\em thorny sphere}. A thorny sphere is a
compact 2-dimensional surface which has $n$ points with conical
singularities, and away from these points is everywhere locally
isometric to a unit sphere $S^2$.
If $l_k$ is a length of a circle of radius $r_k$ around the singular
point $k$, then $\mu_k=2\pi -lim_{r_k\to 0} (l_k/r_k)$ is its angle
deficit.

In section 2 we study thorny spheres isometrically embedded
in a Euclidean 3-space to derive one form of the constraint equation.
In section 3 we show how the thorny sphere can be obtained from a
regular round sphere by set of reconstructions ({\em elementary
deformations}), starting with an arbitrary triangulation of $S^2$.
In section 4 we describe methods of mapping a thorny sphere
onto a unit sphere with cuts.
Constraint equations are derived in section 5
from a consistency condition of these maps and from the
holonomy group of a 3-space that is flat except for the strings.
The special case of small angle deficits
is also considered in this section.
Concrete examples of
thorny spheres with $3$, $4$, and general $n$ conical singularities
are studied in detail in section 6.
Topological aspects of the problem are the subject of section 7.
Finally, in section 8, we demonstrate how
thorny spheres can be used to construct static solutions of the
Einstein equations with $n$ radial cosmic strings.

\section{Thorny spheres embedded in flat Euclidean space
and the number of free parameters}
\setcounter{equation}0

\subsection{Intrinsic geometry of a thorny sphere}

	We shall first consider the large class of thorny spheres
$M^2$ (e.g., all those with three or more conical singularities,
all of which have positive deficit angles)
that are both isometrically embeddable into Euclidean 3-space
(in a unique manner, up to overall translations and rotations)
and also have their geometries uniquely defined by the edge lengths
of geodesic triangulations with the vertices
at the conical singularities.
If $N_f$, $N_e$ and $N_v = n \geq 3$
are the number of triangles, edges and vertices for the
triangulation, then the Euler theorem gives
\begin{equation}\label{0.1}
N_f-N_e+N_v=2\, .
\end{equation}
Since each triangle has 3 edges and each edge belongs to two triangles,
we have $N_e=3N_f/2$.
(Thus the total number of triangles is always even.)
From this relation and the Euler theorem, we get
\begin{equation}\label{0.2}
N_v = n, \hspace{0.5cm} N_e=3n-6\, ,\hspace{0.5cm} N_f=2n-4 \, .
\end{equation}

Except at the conical singularities at the vertices of the triangles,
the thorny sphere has constant Gaussian curvature
$K = {1 \over 2} R$ (with $R$ being the Ricci scalar curvature)
which we shall take to be unity,
and hence each triangle can be isometrically mapped
to a spherical triangle on the unit sphere.

Consider a spherical triangle with vertices 1,2,3. Its edges are parts
of great circles on the sphere. Denote by
$l_1$, $l_2$ and $l_3$  the lengths of the triangle, and by 
$\gamma_1$,  $\gamma_2$ and $\gamma_3$ its interior angles at the
vertices 1,2,3, respectively.  We assume that the edge $l_k$ is
opposite to the $k$th vertex.  For given lengths of the edges, the
angles are uniquely defined,
assuming as we shall that they are all less than $\pi$
(which is indeed the case when the deficit angles are all positive). 
In particular one has
\begin{equation}\label{0.3a}
\cos\gamma_1={\cos l_1 -\cos l_2 \cos l_3
\over \sin l_2 \sin l_3 }\, .
\end{equation}
The triangle can be also specified by its angles. The lengths of the
edges can be determined by using the relation
\begin{equation}\label{0.3}
\cos l_1={\cos \gamma_2 \cos\gamma_3+\cos\gamma_1 
\over \sin\gamma_2 \sin\gamma_3 }\, .
\end{equation}
We also shall use the following expression for the area $A$ of a
spherical triangle:
\begin{equation}\label{0.3b}
A=\gamma_1+\gamma_2+\gamma_3 -\pi\, .
\end{equation}

Because the edges of the triangles are geodesics of the thorny sphere,
adjacent triangles match without producing any singularities
along the edges, except at the vertices.
There one gets a deficit angle $\mu_k$ that is $2\pi$ minus the sum
of the interior angles of the triangles at that vertex.

Thus the entire geometry of each $M^2$ is uniquely determined by
the $N_e=3n-6$ edge lengths of the triangulation,
which can be all specified independently,
within an open set of the $(3n-6)$-parameter space
that is restricted by certain inequalities
(e.g., triangular inequalities).
(For $n=2$, there is no triangulation, and $3n-6 = 0$,
but there is a one-parameter family of thorny spheres
of arbitrary deficit angle $\mu < 2\pi$.  See Appendix B.)

If the deficit angles are all positive, the thorny sphere $M^2$ has
nonnegative Gaussian curvature $K$ everywhere, unit curvature
everywhere away from the conical singularities and Dirac delta-function
curvature at the singularities. The contribution of these two parts of
the curvature to the Gauss-Bonnet theorem is
\begin{equation}
\label{D2c}
 \int_{M^2} K dA = A + \sum_{k=1}^n \mu_k = 4\pi.
\end{equation}

Therefore a thorny sphere can be isometrically rigidly
(i.e., uniquely up to overall translations and rotations)
embedded as a convex surface $M^2$ in 3-dimensional Euclidean space
\cite{Pogo},\cite{Spivak}.

\subsection{Gaussian normal map}

Let ${\bf n}$ be the outward normal
to the embedded surface at each point.
One can then map each point of $M^2$ to a corresponding point
of a unit round $S^2$ also embedded in the 3-dimensional Euclidean space
that has the same unit normal ${\bf n}$.
This map is known as the Gaussian normal map
of the convex surface $M^2$ into $S^2$.
The Gaussian normal map is discussed in more detail in Appendix A,
where it is shown that
\begin{equation}
\label{D3a}
 K \, dA = da,
\end{equation}
where $dA$ is the area element on $M^2$
and $da$ is the corresponding area element on $S^2$
(with the same corresponding set of unit normal vectors ${\bf n}$
in the embedding Euclidean 3-space).

From the divergence theorem applied to the interior of $M^2$,
one can easily prove
\begin{equation}
\label{D3}
 \int_{M^2} {\bf n} \, dA = 0,
\end{equation}
and from the divergence theorem applied to the interior of $S^2$,
one can similarly prove
\begin{equation}
\label{D4}
 \int_{M^2} {\bf n}\, K \, dA = \int_{S^2} {\bf n} \, da = 0.
\end{equation}

The unit normal ${\bf n}$ to the embedding $M^2$ of a
thorny sphere is not well defined at each conical singularity. As one
goes around the $k$th conical singularity infinitesimally close to it,
the unit normal ${\bf n}$ to the embedding (well defined everywhere
except at the conical singularities themselves, and hence defining a
smooth Gaussian normal map from that smooth part of $M^2$ into the
round $S^2$ of unit-normal directions) sweeps out a topological circle
on the round sphere $S^2$ of possible directions for the unit normal,
with the area of the disk $D_k$ within this topological circle on the
round $S^2$ being the deficit angle $\mu_k$. The set of these $n$ disks
is the part of the round $S^2$ that is not mapped into from the smooth
part of the $M^2$ but instead represents the conical singularities.
Thus in the case that $\mu_k$ is not infinitesimal, the direction of
the unit normal at that conical singularity is spread out over this
cone (the disk $D_k$ on the $S^2$) and has an angular uncertainty of
the order of $\mu_k^{1/2}$.

\subsection{Constraint equation}

The integral of Eq. (\ref{D3}) is the same as what it would be
if one excluded the zero-area conical singularities
and inserted the unit curvature $K$ for reg$(M^2)$,
the smooth part of $M^2$:
\begin{equation}
\label{D5}
\int_{{\mbox{\small{reg}}} (M^2)}\, {\bf n}\, K \, dA
  = \int_{S^2-\sum D_k} {\bf n} \, da = 0.
\end{equation}
If one subtracts the second integral here from that of
the second integral of Eq. (\ref{D4}), one gets
\begin{equation}
\label{D6}
 \int_{\sum D_k} {\bf n} \, da = 0,
\end{equation}
which is one version of what we shall call the constraint
equation for the thorny sphere,
restricting the orientation of its conical singularities
in the embedding Euclidean 3-space.

One can regard this constraint as arising from the fact
that we have restricted the thorny sphere to have constant
Gaussian curvature everywhere except at the conical singularities,
and this restricts three combinations of the strengths and
positions of these singularities.
A more physical interpretation of the constraint is as a force-balance
equation, to which we now turn.

One can define
\begin{equation}
\label{D11b}
 {\bf F}_k \equiv \int_{D_k} {\bf n} \, da,
\end{equation}
which can be interpreted as the force,
defined as a vector in the embedding Euclidean 3-space
that can trivially be parallely transported over it
and added to other such forces,
exerted by a string that produces the angle deficit angle $\mu_k$
at the $k$th conical singularity.

	Since the area of the disk $D_k$ is
\begin{equation}
\label{D12}
 \mu_k = \int_{D_k} da
\end{equation}
(see the proof in Appendix A),
one can write
\begin{equation}
\label{D13}
 {\bf F}_k = \mu_k \hat{\bf n}_k,
\end{equation}
where $\hat{\bf n}_k$ is the normal ${\bf n}$ averaged over
the area of the disk $D_k$:
\begin{equation}
\label{D14}
 \hat{\bf n}_k \equiv {\int_{D_k} {\bf n} da \over \int_{D_k} da}.
\end{equation}

For $\sum_{k=1}^n\mu_k \ll 4\pi$, each disk $D_k$ is nearly round, and
the averaged normal will have the length
\begin{equation}
\label{D15}
 |\hat{\bf n}_k| \approx 1 - {\mu_k \over 4\pi},
\end{equation}
with this relation being exact when the disk $D_k$ is precisely round.
Thus the length is nearly unity when $\mu_k$ is small,
but it decreases below that to a minimum value of $1/2$
when $\mu_k$ is increased to its maximum value of $2\pi$
(assuming a round disk $D_k$, which one indeed gets in the case of just
two conical singularities,
a case in which the geometry is not determined by the geodesic
distances between conical singularities, simply $\pi$ in this case,
but which involves an arbitrary deficit angle $\mu < 2\pi$
and is discussed in Appendix B).

For an opposite extreme case, in which $\sum_{k=1}^n\mu_k = 4\pi$
and in which the embedding of the thorny sphere gives the two sides
of a convex polygon, the deficit angle at a vertex is $2\pi$
minus twice the corresponding interior angle of the polygon
(since the surface $M^2$ corresponds to both sides),
and the disk $D_k$ is one interior lune-shaped region between
two great circles that intersect at angle $\mu_k$.
Then one can show that in this extreme case,
\begin{equation}
\label{D16}
 |\hat{\bf n}_k| = {\pi\over\mu_k} \sin{\mu_k \over 4},
\end{equation}
which for small $\mu_k$ has the limit $\pi/4$ rather
than the limit of unity that Eq. (\ref{D15}) has.
Thus for Eq. (\ref{D15}) to be valid, it is required
that the disk $D_k$ be nearly round,
for which it is not sufficient merely
that $\mu_k$ be small, but also that the effects of all the other
conical singularities on the embedding also be small.
	
	In terms of the precisely-defined forces ${\bf F}_k$,
the constraint equation (\ref{D6}) becomes the force-balance
equation that the sum of these forces vanishes:
\begin{equation}
\label{D11}
 {\bf F} \equiv \sum_{k=1}^n {\bf F}_k
 = \sum_{k=1}^n \mu_k\, \hat{{\bf n}}_k
 = \sum_{k=1}^{n} \int_{D_k} {\bf n}\, da = 0.
\end{equation}
This is one precise version of the constraint equation
for arbitrary possible positive deficit angles.

	One way to visualize this constraint equation
is to imagine that one covers the disks $D_k$ of the round $S^2$
with some material with constant mass per unit area.
Then the constraint equation is the condition that the center
of mass of the $n$ disks be at the center of the round sphere.

	Although this form of the constraint equation
is easily visualizable and is
precisely valid for general deficit angles
(so long as they allow the thorny sphere to be rigidly embedded
in Euclidean 3-space, which will be the case
for positive deficit angles but need not be so for
negative deficit angles),
it is not very convenient for calculations,
since for $n > 2$ it is a rather difficult procedure to construct
the embedding of a thorny sphere into Euclidean 3-space.
Therefore, it is also useful to look at other ways of representing
thorny spheres, which we shall do in section 4.

Let us emphasize that the above results can be generalized. Instead of
a thorny sphere one may consider a {\em thornifold}, that is a closed
2-dimensional surface with conical singularities and arbitrary smooth
metric outside them. We assume that this metric has positive Gaussian
curvature $K$ so that it can be isometrically embedded  in  Euclidean
3-space as a closed convex  surface. As it is shown in Appendix A, 
the constraint equation (\ref{D11})  is modified and takes the form
\begin{equation}
\label{D11a}
{\bf F} \equiv \sum_{k=1}^n \mu_k\, \hat{{\bf n}}_k
 = \int_{{\mbox{\small{reg}}} (M^2)}\, {\bf n}\, (1-K)\, dA\, ,
\end{equation}
where $K$ is the Gaussian curvature of $M^2$. We call this relation a
{\em generalized constraint equation}. The presence of non-constant
Gaussian curvature in the right-hand side of  this equation makes
possible the existence of new configurations, e.g.  with a single
conical singularity. See the discussion of $C$-metrics in  Appendix A
for some interesting physical applications of the generalized 
constraint equation.

\section{Spherical triangulations of a thorny sphere}

\setcounter{equation}0

\subsection{Elementary deformation of a sphere
and another count of the degrees of freedom}

In this section we describe how to construct a thorny sphere starting
with a triangulation of a regular unit sphere $S^2$ (see also
\cite{Ivanov}). In the next section we describe the mapping of thorny
spheres into round spheres embedded in three-dimensional Euclidean
space.

We start construction of a thorny sphere by taking a regular unit
sphere with an arbitrary given triangulation of it,
using spherical triangles.
As above, let $N_v = n \geq 3$, $N_e = 3n-6$, and $N_f = 2n-4$
be the number of vertices, edges, and triangles for the triangulation,
with Eqs. (\ref{0.3a}), (\ref{0.3}), and (\ref{0.3b})
applying for the geometry of each triangle.

\begin{figure}
\centerline{\epsfig{file=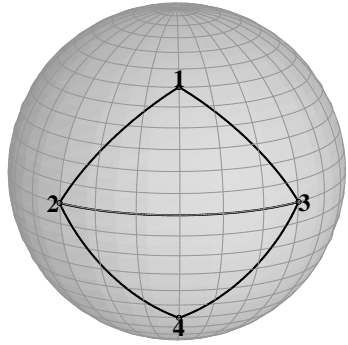, width=4.5cm}}
\vspace{1cm}
\caption[quadra]{Spherical quadrangle composed of two spherical
triangles. The edges are the lines of the large circles of the
sphere.}
\label{quadra}
\end{figure}

Now let us cut from the triangulation a spherical quadrangle $Q$ 
which consists of two triangles with a common edge between
vertices 2 and 3; see Fig.~\ref{quadra}.
We denote the 
length of this common edge by $l_1$. The other edges have lengths 
$l_2$, $l_3$ and $\bar{l}_2$, $\bar{l}_3$ for the first and  second
triangles, respectively. Denote by $Q'$ a new  spherical 
quadrangle for which the length of the common edge is $l'_1$, while 
the other lengths are the same. This change of the length $l_1$
results  in the change of angles of each of two triangles which can be
found by  using (\ref{0.3}). In this procedure the lengths 
$l_2$, $l_3$, $\bar{l}_2$, $\bar{l}_3$ of the
edges of $Q'$  remain the same as for $Q$. That is why the modified
spherical  quadrangle $Q'$ can be glued back into a cut sphere. The
resulting  surface will be smooth everywhere except at 4 vertices
where angle deficits will appear. We call this procedure an {\em
elementary deformation} of  the sphere. 

For a given triangulation,
%in order to create the required angle deficits in $N_v = n$ vertices,
one can use $N_e = 3n-6$ elementary deformations independently
to fix all of the edge lengths (and hence also all the deficit angles)
and thus to determine the $(3n-6)$-parameter metric
on a generic thorny sphere with $n$ conical singularities.
However, if we have the goal of fixing required deficit angles
at the $N_v = n$ vertices, not all of the elementary deformations
have independent effects upon them.
Some combinations of these  deformations do not generate angle deficits,
but simply move vertices of the triangulation along the sphere.
The number of such `degrees of freedom'
that do not affect the deficit angles
is $2N_v - 3 = 2n-3$ (two `degrees of freedom'
per vertex minus 3 `degrees of freedom' corresponding to rigid
rotations of the sphere which  preserve the lengths of each edge
unchanged). Thus the total number of `real degrees of freedom' which
generate deformations in the angle deficits is $N_e-2N_v+3= n-3$. These
deformations are sufficient to create the required angle deficits
at all the $n$ vertices except 3.
This is exactly what one can expect in a
general case, since there exist exactly 3 consistency conditions
in the vector constraint equations (\ref{i0}) and (\ref{i1})
relating angle deficits and positions of the singular
points on the thorny sphere.

The above counting of the `degrees of freedom' gives us also the 
following useful information. Let us fix the values of all $n$ angle 
deficits $\mu_k$. Then $2n$ `degrees of freedom' characterizing the 
positions of the vertices must obey 3 constraint equations. We can also 
use 3 `degrees of freedom' of the rigid rotation of the sphere to 
put, say, the first point to the north pole of the sphere, and the 
second one on the meridian $\phi=0$. After this there remains 
$2n-6$ free parameters. By adding to them the $n$ parameters
$\mu_k$ we return to the $3n-6$ parameters of a generic
thorny sphere with $n$ conical singularities.

\subsection{Constraints for elementary deformations of a sphere}

Now to illustrate the nature of these constraints we discuss
a case when all angle deficits are infinitesimally small. 
Consider  an infinitesimal elementary deformation of a spherical 
quadrangle $Q$ shown on Fig. \ref{quadra}.
To produce conical singularities one can deform the length 
of the common edge $(2,3)$
while keeping the lengths of the other edges unchanged. This deformation
changes internal angles at vertices 1,2,3,4 and yields conical
singularities. Let us introduce the following notations
\begin{equation}\label{16}
w_3=({\bf n}_2 \cdot {\bf n}_1),~~w_2=({\bf n}_3 \cdot {\bf n}_1),~~
\bar{w}_3=({\bf n}_2 \cdot {\bf n}_4),~~
\bar{w}_2=({\bf n}_3 \cdot {\bf n}_4),~~
w_1=({\bf n}_2 \cdot {\bf n}_3),
\end{equation}
where ${\bf n}_k$ are the unit vectors (with the beginning at the center
of the sphere)
which define the positions
of the corresponding vertices. 
If $l_k$ is the length of the
edge opposite to the $k$th vertex, then $w_k=\cos l_k$, which is
true for triangles whose edges are arcs of great circles.
For triangles on $S^2$, Eq. (\ref{0.3}) gives,
\begin{equation}\label{17}
w_1={c_2c_3+c_1 \over s_2 s_3},~~
w_2={c_1c_3+c_2 \over s_1 s_3},~~
w_3={c_2c_1+c_3 \over s_2 s_1},
\end{equation}
where $c_k=\cos ~\gamma_k$, $s_k=\sin ~\gamma_k$. Analogous
relations for $\bar{w}_k$ can be obtained from (\ref{17})
by replacing $\gamma_k$ by $\bar{\gamma}_k$.
Variations of $w_1$ produce changes of the angles 
$\delta \gamma_k\equiv -x_k$, $\delta \bar{\gamma}_k\equiv -
\bar{x}_k$.  The condition that these variations do not change
$w_2$ can be written in linear order as
$$
\delta w_2={1 \over s_1s_3}\left((c_1s_3 w_2+c_3s_1)x_1+
(c_3s_1 w_2+c_1s_3)x_3+x_2s_2\right)
$$
\begin{equation}\label{18}
={s_2 \over s_1s_3}(w_3x_1+w_1x_3+x_2)=0.
\end{equation}
To get the last line we used (\ref{17}).  A similar
relation follows from the variation of $\bar{w}_2$. Their combination
yields 
$$
(w_3x_1+w_1x_3+x_2)+(\bar{w}_3\bar{x}_1+\bar{w}_1\bar{x}_3+\bar{x}_2)
=w_3 x_1+(x_2+\bar{x}_2)+w_1(x_3+\bar{x}_3)+\bar{w}_3\bar{x}_1
$$
\begin{equation}\label{19}
=w_3\mu_1+\mu_2+\mu_3 w_1+\mu_4\bar{w}_3
=0,
\end{equation}
where we took  into account that $w_1=\bar{w}_1$.
It is easy to see that
$\mu_1=x_1, \ \mu_2=x_2+\bar{x}_2, \   
\mu_3=x_3+\bar{x}_3$, and $\mu_4=\bar{x}_1$
are the conical angle deficits produced by the deformation at 
the vertices $1, 2, 3, 4$.
Consider the particular form of (\ref{i1})
when only these four vertices have nonzero deficit angles:
\begin{equation}\label{20}
{\bf n}_1\mu_1+{\bf n}_2\mu_2+{\bf n}_3\mu_3 +{\bf n}_4\mu_4
=0.
\end{equation}
Under projecting it on the vector ${\bf n}_2$
we get exactly (\ref{19}).

	To get Eq. (\ref{20}) directly, and not just its projection
on ${\bf n}_2$, we can do some further algebra and find that
for each of the two spherical triangles with only the length $l_1$
perturbed,
\begin{equation}\label{20b}
x_1{\bf n}_1 + x_2{\bf n}_2 + x_3{\bf n}_3
 = {\delta w_1 \over 1-w_1^2}\,{\bf n}_3\times{\bf n}_2,
\end{equation}
\begin{equation}\label{20c}
\bar{x}_1{\bf n}_4 + \bar{x}_2{\bf n}_2 + \bar{x}_3{\bf n}_3
 = {\delta w_1 \over 1-w_1^2}\,{\bf n}_2\times{\bf n}_3.
\end{equation}
When these two equations are added, the right hand sides cancel,
and one gets Eq. (\ref{20}).

	A generalization of this linearized vector condition to
finite deformations will be given in the next sections.

\section{Mapping a thorny sphere onto a round sphere with cuts}

As we already mentioned, there exists an  embedding of a thorny sphere
in flat space (at least for positive deficit angles). 
But practically it is very difficult to obtain this
embedding explicitly and get the precisely-defined forces ${\bf F}_k$
for the force balance equation (\ref{D11}) unless the angle deficits
are small.
An exception for large deficits 
is a simple case when a thorny sphere has two conical singularities,
see Appendix B.
(This case is not covered by Section 2 since it cannot be
triangulated using as vertices only the two conical singularities).
Therefore, in this section we present another
description of a thorny sphere by mapping it onto a round
sphere with cuts. This approach allows us to formulate the constraint
equation in an {\em explicit} algebraic form. 
There are several ways to do this. We describe here two simple methods.

\subsection{Method A}
\setcounter{equation}0

One method of representing the thorny sphere,
$\tilde{S}^2$, with $n = N_v$ conical singularities
$A_k$, $1 \leq k \leq n$,
is the following, applicable for $n \geq 3$:
Let the singularities be labeled so that the sequence of
the shortest geodesic
from $A_1$ to $A_2$
(say geodesic segment $\gamma_1$ with beginning at $A_1$ and end at
$A_2$), that from $A_2$ to $A_3$
(say $\gamma_2$), \ldots, that from $A_{(n-1)}$ to $A_n$
(say $\gamma_{(n-1)}$), and that from $A_n$ to $A_1$
(say $\gamma_n$) forms a closed path $\tilde{\cal P}$
that does not intersect itself.
For example, one can choose some regular point,
find the shortest geodesic from that point to
each conical singularity,
arbitrarily choose one of the singular points to be $A_1$,
and then label the remaining conical
singularities in the same order as the angles, at the regular point,
of the tangent vectors of these geodesics from the regular point
to those singularities.
It is not obvious whether or not the resulting sequence of geodesics
between the conical singularities chosen in this order
(or in any other possible specification of the order)
will necessarily form a closed path $\tilde{\cal P}$
that does not intersect itself, but here we shall assume that it does.

\begin{figure}
\centerline{\epsfig{file=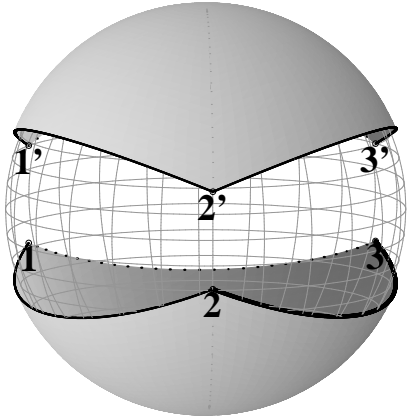, height=5.5cm}}
\caption[orange_2]{A unit sphere with cuts. By gluing the cuts
one obtains a sphere with three conical singularities.}
\label{orange_2}
\end{figure}

	Now the closed path ${\cal P}$ divides
the thorny sphere, $\tilde{S}^2$, into two parts,
say ${\tilde{\cal M}}_1$, which is encircled clockwise
by ${\cal P}$,
and ${\tilde{\cal M}}_2$, which is encircled counterclockwise
by ${\cal P}$.
Because the interiors of both contains no conical singularities,
they can be isometrically mapped to corresponding regions,
${\cal M}_1$ and ${\cal M}_2$, on the round unit $S^2$
of the same unit curvature as the part of the thorny sphere
$\tilde{S}^2$ away from the conical singularities.
The boundaries of these two regions on $S^2$ are the
geodesic polygons that are the images
${\cal B}_1$ and ${\cal B}_2$ of $\tilde{\cal P}$ in these maps
from the thorny sphere $\tilde{S}^2$ into the round sphere $S^2$.
In the simplest case of three conical singularities the 
regions $\tilde{\cal M}_1$ and $\tilde{\cal M}_2$
are spherical triangles. Their map on the sphere is shown on
Fig. \ref{orange_2}.

	The two maps preserve the lengths of the geodesic
edges $\gamma_k$ of the polygon,
and they preserve the angles $\phi_{(1,k)}$ and $\phi_{(2,k)}$
between the two successive
geodesics that meet at the conical singularity $A_k$.
Let us define $\phi_{(1,k)}$ to be the angle between
the tangent vector of the
geodesic ending at $A_k$ and that of the geodesic beginning
at $A_k$, measured in the region ${\tilde{\cal M}}_1$
and taken to be positive if clockwise,
so that the interior angle at that vertex of the polygon
is $\pi - \phi_{(1,k)}$.
Similarly, define $\phi_{(2,k)}$ to be the angle between that the
geodesic ending at $A_k$ and that of the geodesic beginning
at $A_k$, measured in the region ${\tilde{\cal M}}_2$
and also taken to be positive if clockwise,
so that the interior angle at that vertex of the polygon
is $\pi + \phi_{(2,k)}$ (now with a plus sign
since with the ordering given for the geodesic edges,
the polygon encircles ${\tilde{\cal M}}_2$
in the counterclockwise orientation rather
than in the clockwise orientation as it does ${\tilde{\cal M}}_1$).
Then the conical deficit angle at $A_k$ is
$\mu_k = \phi_{(1,k)} - \phi_{(2,k)}$.

	The maps from ${\tilde{\cal M}}_1$ on the thorny sphere
$\tilde{S}^2$ to ${\cal M}_1$ on the round sphere $S^2$,
and from ${\tilde{\cal M}}_2$ on $\tilde{S}^2$ to ${\cal M}_2$
on $S^2$, give points $A_{(1,k)}$ on $S^2$ that are the vertices
of ${\cal M}_1$, and points $A_{(2,k)}$ on $S^2$ that are the vertices
of ${\cal M}_2$.  Then the locations of the $2n$ points
$A_{(1,k)}$ and $A_{(2,k)}$ on the round sphere $S^2$
uniquely determine the geometry of the thorny sphere $\tilde{S}^2$,
since they determine the polygon boundaries
${\cal B}_1$ and ${\cal B}_2$ of the regions
${\cal M}_1$ and ${\cal M}_2$ whose interiors have the unit-curvature
metric inherited from the unit sphere $S^2$.  Because the boundary
segments are geodesics, the successive ones from
${\cal M}_1$ and ${\cal M}_2$ can be identified so that the union
of the two regions with this identification forms the thorny
sphere $\tilde{S}^2$ with no singularities except for the
conical singularities with deficit angles $\mu_k$ at the $n$ vertices.

\begin{figure}
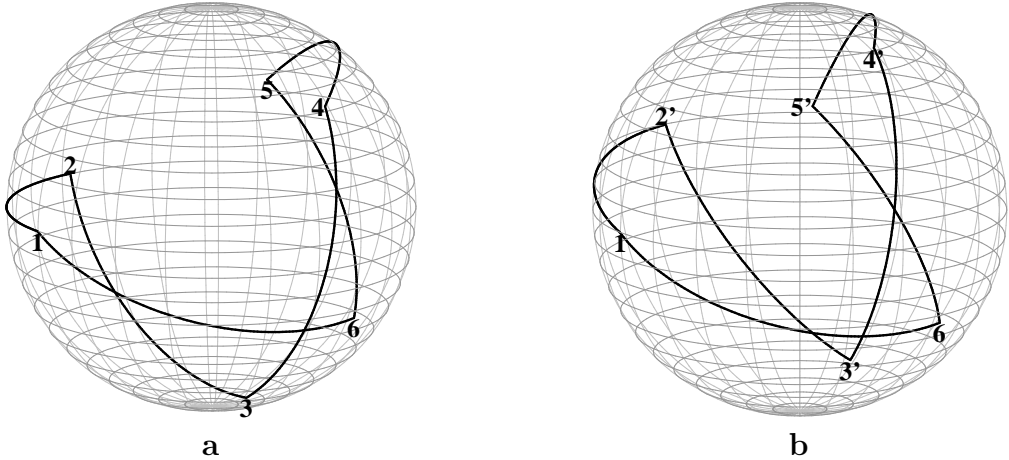

\begin{center}
\[
\begin{array}{cc}
\hspace{-1cm}
\epsfig{file=don-201.eps, height=5.5cm}
&
\hspace{2cm}\epsfig{file=don-301.eps, height=5.5cm}\\
\hspace{-1cm}{\bf a} & \hspace{2cm}{\bf b}
\end{array}
\]
\end{center}
\caption[f1]{Figure (a) demonstrates the region ${\cal M}_1$
for a the thorny sphere with 6 singularities.
${\cal M}_1$ lies on the right
when one goes from point 1 to point 2.
Figure (b) shows region ${\cal M}_2$ for the same sphere.
It lies on the left hand side
when one goes from point 1 to point 2'.
On the both pictures 
points 1,5,5' and 6 lie on the back side of the sphere
}
\label{A1}
\end{figure}

\begin{figure}
\centerline{\epsfig{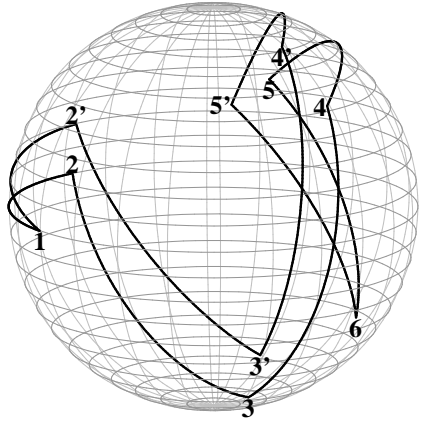}}
\caption[f1]{This figure demonstrates the region ${\cal M}_C$
constructed by method A.
Points 1,5,5' and 6 lie on the back side of the sphere.
Region ${\cal M}_C$ is the unification of regions
${\cal M}_1$ and ${\cal M}_2$ shown on Fig. \ref{A1}
The thorny sphere is obtained by gluing points $k$ and $k'$.
}
\label{A2}
\end{figure}

	Let us see how we get the right parameter count from
this construction.  Na\"{\i}vely one has two parameters for locating
each of the $2n$ points $A_{(1,k)}$ and $A_{(2,k)}$
on the round sphere $S^2$, or $4n$ total.  However, there is the
constraint that the geodesics segments from the successive
$A_{(1,k)}$'s must match those from the successive $A_{(2,k)}$'s,
which gives $n$ constraints, leaving only $3n$ parameters arbitrary.
Then there is an arbitrary 3-parameter
rotation that one can separately apply to both ${\cal M}_1$
and ${\cal M}_2$, so of the $3n$ arbitrary parameters,
only $3n-6$ are physically significant in determining the geometry
of $\tilde{S}^2$.  This is precisely
equal to the number of continuous parameters of a thorny
sphere with $n$ conical singularities of arbitrary strength,
the number of edge lengths $N_e$ of a triangulation of it,
as discussed above (under the assumption that the triangulation
exists, though the number of continuous parameters
would not be expected to depend on this assumption).

	Given the location of ${\cal M}_1$ on the round sphere
$S^2$ (which has three Euler-angle parameters of arbitrariness),
one could fix the three arbitrary rotation angles for the location
of ${\cal M}_2$ on the $S^2$ so that two successive vertices
of ${\cal M}_1$
(say $A_{(1,n)}$ and $A_{(1,1)}$ in order to refer to them
explicitly below)
coincide with the corresponding ones of ${\cal M}_2$
(i.e., $A_{(2,n)}$ and $A_{(2,1)}$ respectively),
with these two regions being on opposite sides of the geodesic segment
$\gamma_n$ joining these two successive vertices
and providing the common boundary between ${\cal M}_1$ and ${\cal M}_2$.
The union of these two regions with their common boundary, $\gamma_n$,
no longer a boundary, then gives one single simply-connected region
on the round $S^2$ that represents the thorny sphere $\tilde{S}^2$.
We will denote this union region as ${\cal M}_C$,
and its boundary as $C$.
Note, however, it is not ensured that the two regions that have been
joined, ${\cal M}_1$ and ${\cal M}_2$, will not overlap somewhere
other than where they have been joined,
so the map from $\tilde{S}^2$ to its image in $S^2$
is not necessarily one-to-one but can in some regions be two-to-one,
and in general $C$ may have self-intersections.

In Fig. \ref{A1} we demonstrate the regions for $n=6$ 
conical singularities. Regions  ${\cal M}_1$ and ${\cal M}_2$
are shown on pictures (a) and (b), respectively. 
Points 1 and 6 have the same coordinates for both regions.
In Fig. \ref{A2} ${\cal M}_1$ and ${\cal M}_2$ are united
in the region ${\cal M}_C$ by gluing them along the edge 
between points 1 and 6. The closed curve on 
Fig. \ref{A2} is the boundary $C$ of ${\cal M}_C$, and in the considered
example it does not have self intersections. Hence, one can cut
the region inside $C$ and then glue the corresponding points
(2 with $2'$, 3 with $3'$, 4 with $4'$ and 5 with $5'$).
This yields a thorny sphere with some configuration of 6 
conical singularities with deficit angle $\mu=\pi/12$.

The boundary $C$ has $2n-2$ vertices
$A_{(1,1)} = A_{(2,1)}$, $A_{(1,n)} = A_{(2,n)}$, $A_{(1,k)}$
for $2 \leq k \leq n-1$, and $A_{(2,k)}$ for $2 \leq k \leq n-1$,
which can be at arbitrary locations except for the constraints
that the $n-1$ geodesic distances between the successive
$A_{(1,k)}$'s must be the same as those between the corresponding
successive $A_{(2,k)}$'s ($n-1$ constraints, since we have already
imposed the fact that the distance between $A_{(1,1)}$ and $A_{(1,n)}$
is the same as that between $A_{(2,1)}$ and $A_{(2,n)}$
by putting $A_{(1,1)}$ at the same location as $A_{(2,1)}$
and $A_{(1,n)}$ at the same location as $A_{(2,n)}$).
Therefore, the number of free parameters done this way is
twice the $2n-2$ vertices, minus the $n-1$ constraints,
or $3n-3$, all of which are arbitrary, but of which three merely
determine the orientation of the entire region on the $S^2$,
so that the number of true physical parameters is $3n-6$.
(One could arbitrarily remove these three remaining gauge parameters
of the freedom to rotate the coordinates
by putting, say, $A_{(1,1)}$ at the ``north pole'' of the $S^2$,
at polar angle $\theta = 0$, and then putting $A_{(1,n)}$
along the ``prime meridian,'' $\phi = 0$.)

	As we shall see below, the $3n-6$ coordinate-independent
parameters that are in one-to-one correspondence with
the geometries on a unit-curvature thorny sphere with $n$
conical singularities (up to the discrete choice of the ordering
of the singularities in the construction above, or of the
choice of the triangulation when one takes its edge lengths
as the $3n-6$ parameters)
can be interpreted as $2n-3$ parameters for
the relative locations of the $n$ conical singularities on some $S^2$
(i.e., after taking out an overall rotation),
plus $n-3$ conical deficit angles that can be freely specified
once the relative locations are fixed.
There is then a constraint fixing three of the conical deficit
angles, which, at least in the case of small deficit angles,
becomes the force-balance Eq. (\ref{i1}) for the strings
that produce the deficit angles.

	One way to see this constraint on the deficit angles
is to consider how much freedom one has
to specify the deficit angles $\mu_k$ after the points $A_{(1,k)}$
have been specified.  Specifying the points $A_{(1,k)}$
determines the angles $\phi_{(1,k)}$ between the successive
geodesics joining those points, but the deficit angles are
$\mu_k = \phi_{(1,k)} - \phi_{(2,k)}$, and the angles $\phi_{(2,k)}$
are determined by the location of the points $A_{(2,k)}$.
As noted above, without loss of generality we can orient ${\cal M}_2$
relative to ${\cal M}_1$ so that $A_{(1,1)}$ coincides with $A_{(2,1)}$
and $A_{(1,n)}$ coincides with $A_{(2,n)}$.
Then the successive $A_{(2,k+1)}$'s must be at the same distances
from the $A_{(2,k)}$'s as the $A_{(1,k+1)}$'s are from the
$A_{(1,k)}$'s, but for $k < n-2$, the direction from $A_{(2,k)}$
to $A_{(2,k+1)}$ (at angle $\phi_{(2,k)}$ clockwise from the direction
of the geodesic coming from $A_{(2,k-1)}$ to $A_{(2,k)}$)
is a free parameter, whose choice fixes the deficit angle
$\mu_k = \phi_{(1,k)} - \phi_{(2,k)}$ at that vertex.
However, when one gets to $A_{(2,n-2)}$, the angle $\phi_{(2,n-2)}$
is fixed (up to a two-fold degeneracy) so that the vertex
$A_{(2,n-1)}$ is at the same distance from $A_{(2,n)}$
as $A_{(1,n-1)}$ is from $A_{(1,n)}$.  Then when $A_{(2,n-1)}$
is thus fixed, the angles $\phi_{(2,n-2)}$, $\phi_{(2,n-1)}$,
and $\phi_{(2,n)}$ are fixed, and hence these final three
deficit angles, $\mu_{(n-2)}$, $\mu_{(n-1)}$, and $\mu_n$,
are determined
(up to the two-fold degeneracy of the two possible locations
for $A_{(2,n-1)}$ at the fixed distances from $A_{(2,n-2)}$
and from $A_{(2,n)}$) and are not free parameters.
Therefore, once the $2n-3$ parameters of the relative locations
$A_{(1,k)}$ of the $n$
conical singularities are determined (as seen from within ${\cal M}_1$),
one is free to specify only $n-3$ of the deficit angles,
giving a total of $3n-6$ parameters.

	Another way to express the three constraints on the
deficit angles once their $A_{(1,k)}$ locations are fixed is
from the constraint on the holonomy from going successively
around all of the conical singularities in a three-dimensional space.
This will be discussed in Section 5.

\subsection{Method B}
\setcounter{equation}0

Another possible method to cut the thorny sphere 
with conical singularities into a piece that can be fit onto
the round sphere is the following:
Let us choose one of the singular points, say point $A_n$, 
and connect it
by shortest geodesics with the rest of the points,
$A_k$, $k=1,\ldots, n-1$.
Since the thorny sphere is a complete Riemannian manifold,
the shortest geodesics which connect
$A_n$ with different $A_k$ do not intersect.
If there are more than one
geodesic with the same length, we choose one of them.
Let us also make the following convention for 
ordering
of points $A_k$,  $k=1,..., n-1$. 
Choose one of these points and denote it by $A_1$.
We can go clockwise
around $A_n$ starting from the geodesic between $A_n$ and $A_1$.
The convention is that the next geodesic we hit 
corresponds to the geodesic between $A_n$ and $A_2$,
the next geodesic after that corresponds to
point $A_3$, etc.

\begin{figure}
\centerline{\epsfig{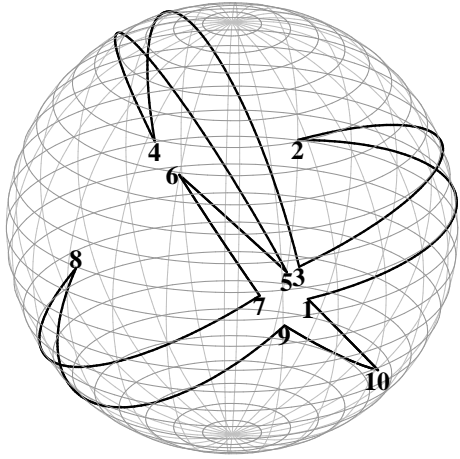}}
\caption[f1]{This figure demonstrates the method B.
Points 2,4,8 on the figure lie on the back side of the sphere.
After cutting the star-like region and gluing the rest part 
one get a sphere with six conical singularities with deficit angle 
$\mu=\pi/12$. After gluing points 1,3,5,7 and 9 are identified.}
\label{MB}
\end{figure}

Let us now make cuts of $\tilde{S}^2$ from $A_n$ to points
$A_k$ along the geodesics. This procedure yields a spherical 
polygon ${\cal M}_C$ with a boundary $C$ which is a closed
curve. 
In general, this polygon will have a shape different
from the polygon obtained by Method A,
although the number of edges and vertices is the same, $2n-2$.
The boundary now consists of $n-1$ pairs of geodesics with
equal lengths. The geodesics in the given pair are connected 
at a vertex which corresponds to point $A_n$ on $\tilde{S}^2$, 
while different 
pairs are connected at vertices $A_k$, $k=1,...,n-1$. 
Let denote by $B_{2k}$ the vertex
which corresponds to $A_k$, $k=1,...,n-1$. 
Then a vertex between $B_{2(k-1)}$ and $B_{2k}$ is the image of $A_n$. 
We denote it by $B_{2k-1}$.

As in the previous case ${\cal M}_C$ can be glued on 
a regular sphere $S^2$, and
the boundary $C$ of ${\cal M}_C$ will be mapped on a closed 
contour on $S^2$.
Each vertex $B_k$ on $C$ has a uniquely
defined coordinate ${\bf x}_k$ on $S^2$.  
We can now  define coordinates of the singular
points on $\tilde{S}^2$ as follows: points $A_k$ ($k=1,..,n-1$)
have coordinates ${\bf n}_k={\bf x}_{2k}$ (coordinates of $B_{2k}$),
and the coordinate ${\bf n}_n$ of $A_n$ can be chosen as the coordinate
of one of $B_{2k+1}$. It is convenient to put
${\bf n}_n={\bf x_1}$.

The internal angles at points $B_{2k}$ are $\beta_{2k}=
\alpha_k=2\pi-\mu_k$, which are polar angles around $A_k$, $k=1,...n-1$,
and $\mu_k$ are conical angle deficits at $A_k$.
If $\beta_{2k+1}$ are internal angles at points $B_{2k+1}$, then
they are related to the polar angle around $A_n$ as
\begin{equation}\label{1ab}
\alpha_n=\sum_{k=0}^{n-2}\beta_{2k+1}.
\end{equation}
By assumption, all $\mu_k>0$ and hence $0<\alpha_k<2\pi$,
$k=1,...,n$.
The remarkable property  of the contour $C$ is that
the edge between $B_{2k+1}$ and $B_{2k}$ can be obtained 
by rotating the edge between  $B_{2k-1}$ and $B_{2k}$ 
counterclockwise around  $B_{2k}$ by angle $\beta_{2k}$.
An example of ${\cal M}_C$ obtained by
cutting the sphere by the given method is shown on Fig. \ref{MB}.

\section{Constraint equation}
\setcounter{equation}0

\subsection{Derivation of the constraint equation}

The procedure to construct a sphere with conical singularities
either by method A or B requires that the contour $C$ be closed,
see Figures \ref{A2} and \ref{MB}.
This imposes a constraint on the positions of the vertices on $C$.
We describe how this restriction can be found by using method
A of cutting the sphere. One can show that constraint 
required for method B is the same. 

Consider a region ${\cal M}_C$ which appears after cutting $\tilde{S}^2$
by method A.
Denote coordinates of points $A_{(1,k)}$ on $S^2$ by ${\bf n}_k$,
$k=1,..,n$
and coordinates of points $A_{(2,k)}$ by ${\bf n}'_k$, $k=2,..,n-1$.
Let us also introduce matrices $O({\bf n},\alpha)$ which
belong to the group $SO(3)$ and describe rotations by angle
$\alpha$ around the axes defined by the unit vector ${\bf n}$
normal to the $S^2$.
Our convention is that positive $\alpha$ corresponds to
%under the rotation vector $\bf n$ remains on the left
counterclockwise rotation around $\bf n$
(as seen by looking down upon the sphere, with $\bf n$ up).
Define matrices $O_k=O({\bf n}_k,\mu_k)$, $k=1,..,n$,
where $\mu_k$ is the conical angle deficit at the corresponding 
singular point 
on $\tilde{S}^2$. Given matrices $O_k$ and coordinates
of the vertices $A_{(1,k)}$ on the boundary $C$, coordinates of
the rest of the vertices $A_{(2,k)}$ on $C$ can be found as follows:

The angle at the vertex $A_{(1,1)}$ between geodesics connecting
$A_{(1,1)}$ with $A_{(1,2)}$  and $A_{(2,2)}$ is $\mu_1$. Therefore,
$A_{(2,2)}$ is the image of $A_{(1,2)}$ obtained by rotating
$A_{(1,2)}$ around $A_{(1,1)}$ by angle $\mu_1$,
and by using the matrices one can write  
\begin{equation}\label{n1}
{\bf n}_2'=O_1~{\bf n}_2.
\end{equation} 
Consider now a rotation of $A_{(1,3)}$ around $A_{(1,2)}$
by angle $\mu_2$.
This gives a point $A'_3$ which 
could be obtained by making a cut on thorny sphere
which goes through  
$A_{(1,2)}$ and $A_{(1,3)}$.
To get $A_{(2,3)}$ one has to do an additional rotation of $A'_3$
around $A_{(1,1)}$ by angle $\mu_1$.
This second  rotation  takes into account
that point $A_{(1,2)}$ itself has to be rotated around  $A_{(1,1)}$.
Thus, for coordinates of $A_{(2,3)}$ one gets
\begin{equation}\label{n2}
{\bf n}_3'=O_1~O_2~{\bf n}_3.
\end{equation}
This procedure can be continued further to get coordinates
of the other points $A_{(2,k)}$ 
\begin{equation}\label{n3}
{\bf n}_k'=O_1~O_2\cdot\cdot \cdot O_{k-1}~{\bf n}_k.
\end{equation}
If $k=n$ we come to the final point
$A_{(1,n)}$. This point and its image coincide,
\begin{equation}\label{n4}
{\bf n}_n'={\bf n}_n.
\end{equation}
This means that the vector ${\bf n}_n$
is on the axis of rotation defined by the matrix
\begin{equation}\label{n5}
M=O_1~O_2\cdot\cdot \cdot O_{n-1}.
\end{equation}
So for $M$ we can write $M=O({\bf n}_n,\alpha)$
where $\alpha$ is some angle. 

We can also construct coordinates of the images in a different way
by starting with the point $A_{(2,n-1)}$ which is obtained
by the rotation of $A_{(1,n-1)}$
around $A_{(1,n)}$ by the angle $\mu_n$.
According with our convention, this rotation should be in the opposite 
direction, so for coordinates of the points we get
\begin{equation}\label{n6}
{\bf n}_{n-1}'=O_{n}^{-1}~{\bf n}_{n-1}.
\end{equation}
By  proceeding as earlier we get for the coordinates 
of the $k$th point
\begin{equation}\label{n7}
{\bf n}_{k}'=O_{n}^{-1}~O_{n-1}^{-1}\cdot \cdot
 \cdot O_{k+1}^{-1}~{\bf n}_{k}.
\end{equation} 
Because ${\bf n}_1$ and its image coincide the matrix
\begin{equation}\label{n8}
N=O_{n}^{-1}~O_{n-1}^{-1}\cdot \cdot \cdot O_{2}^{-1}
\end{equation} 
corresponds to a rotation around ${\bf n}_1$ by some angle $\beta$,
$N=O({\bf n}_1,\beta)$.
By using (\ref{n5}) and (\ref{n8}) we can write
\begin{equation}\label{n9}
M~O_{n}=O({\bf n}_n,\mu+\alpha),~~~O_1~N^{-1}=O({\bf n}_1,\mu-\beta).
\end{equation} 
This relation shows that matrices $O({\bf n}_n,\mu+\alpha)$ 
and $O({\bf n}_1,\mu-\beta)$ are identical. In the general case,
if points $A_{(1,1)}$, $A_{(1,n)}$ are not on the same axis,
this implies that the matrices are the unit matrices.
Thus, we come to the following
constraint equation which follows from (\ref{n9})
\begin{equation}\label{n10}
O({\bf n}_1,\mu_1)~O({\bf n}_2,\mu_2)\cdot\cdot \cdot O({\bf n}_n,\mu_n)
=I_3,
\end{equation}
where $I_3$ is the unit 3 by 3 matrix. 

Fig. \ref{A1} gives an example in which the coordinates of the points
on (a) obey the
condition (\ref{n10}) for $\mu=\pi/12$.  The method of rotations 
described above was used
to produce images  $2',3',4',5'$ of points $2,3,4,5$, respectively. 
The constraint equation is also obeyed
for the 6 points $2,4,6,8,10,1$ in Fig. \ref{MB} with the same angle
deficit $\mu=\pi/12$.

\subsection{Alternate derivation of the constraint equation
using holonomies}

	Let the 2-metric on the thorny sphere be $d{\tilde\Omega}^2$
(with unit Gaussian curvature away from the conical singularities),
and consider the following three dimensional metric that is flat
everywhere away from the conical singularities,
which form strings in the radial directions:
\begin{equation}
\label{A}
 ds^2 = dr^2 + r^2 d{\tilde\Omega}^2.
\end{equation}
One can then calculate the holonomy of going around various closed
curves and parallel transporting one's frame.
Since the space is flat except at the strings,
the holonomy will be trivial when the closed curve can be shrunk
to a point without crossing any strings, but it will
generally be nontrivial when the closed curve encircles one
or more strings.

	Choose a regular point, say $A_0$ in ${\tilde{\cal M}}_1$, and
take a curve that starts at $A_0$ and stays in ${\tilde{\cal M}}_1$
until it nears the conical singularity at $A_{(1,k)}$. Then have the
curve encircle that singularity (but no other one), in the 
counterclockwise
direction (by going briefly into ${\tilde{\cal M}}_2$), and then
return in ${\tilde{\cal M}}_1$ to $A_0$. Let the element of the
holonomy group generated by that curve be labeled $H_k$.

	All of the elements of the holonomy group can be obtained
by products of these elements and their inverses.  For example,
the curve that first goes out from $A_0$ to encircle $A_{(1,k)}$
clockwise and then goes across ${\tilde{\cal M}}_1$ to encircle
$A_{(1,j)}$ clockwise before returning in ${\tilde{\cal M}}_1$
to $A_0$ is $H_j^{-1} H_k^{-1}$.  To take a slightly more complicated
example, the curve that goes out from $A_0$ to leave
${\tilde{\cal M}}_1$ between $A_{(1,1)}$ and $A_{(1,2)}$
and then goes across ${\tilde{\cal M}}_2$ to encircle
$A_{(2,4)}$ clockwise and then return back along its previous 
path to $A_0$ generates the holonomy group element
$\tilde{H}_4=H_2 H_3 H_4^{-1} H_3^{-1} H_2^{-1}$.

	Now consider the curve that starts at $A_0$, goes out
in ${\tilde{\cal M}}_1$ and encircles $A_{(1,n)}$ counterclockwise,
returns directly in ${\tilde{\cal M}}_1$ to $A_0$,
and then in turn goes out and encircles $A_{(1,n-1)}$ 
counterclockwise
and returns, and then encircles $A_{(1,n-2)}$ counterclockwise, etc.,
until finally it encircles $A_{(1,n)}$ counterclockwise and returns 
to $A_0$.
Thus it encircles each of the conical singularities clockwise
in order, staying in ${\tilde{\cal M}}_1$ except for each
time it encircles a singularity.  The total holonomy generated
by this curve is $H_1 H_2 H_3 \cdots H_{n-1} H_n$.

	But since this curve encircles all of the singularities, it can
be deformed so that all of it lies in ${\tilde{\cal M}}_2$ except
for the initial part leaving $A_0$ and the final part returning
to $A_0$.  This curve can then be shrunk to zero without crossing
any singularities, so for consistency it must represent the trivial
holonomy element (the identity).
Therefore, we get the constraint equation
\begin{equation}
\label{B}
H_1 H_2 H_3 \cdots H_{n-1} H_n = I.
\end{equation}
In the $SO(3)$ representation this constraint coincides
with equation (\ref{n10}) which was found by the alternate computation,
with $H_k=O({\bf n}_k,\mu_k)$.

	If we use an $SU(2)$ spinor representation of the holonomy,
then the conical singularity at $A_1$ with deficit angle $\mu_k$
generates the holonomy element
\begin{equation}
\label{C}
 H_k = U_k = U({\bf n}_k,\mu_k) \equiv e^{\frac i2 \not{n}_k \mu_k}
     = I\cos{\frac {\mu_k}2}+i{{\not\!\! n}_k}~\sin{\frac {\mu_k}2} ,
\end{equation}
where
\begin{equation}
\label{Dba}
\not\!\!{n}_k \equiv {\bf n}_k \cdot \sigma \equiv n^i_k \sigma_i
 = \left( \begin{array}{lr}
 n_k^3        & n_k^1 + in_k^2 \\
 n_k^1 - in_k^2 &      - n_k^3
 \end{array} \right)
\end{equation}
with $n_k^i$ being the $i$-th Cartesian
coordinates of the unit normal $\bf{n}_k$
to the round $S^2$ embedded in flat
three-space, at the point $A_{(1,k)}$ of ${\cal M}_1$ that
represents $A_k$ on the thorny sphere $\tilde{S}^2$.
The Pauli matrices $\sigma_i$ are chosen such that 
$i\sigma_1\sigma_2\sigma_3=1$ which guarantees that
$U_k$ corresponds to counterclockwise rotation
around ${\bf n}_k$.
For generic assumed deficit angles, without imposing
the constraint (\ref{B}), the product of all the $U_k$'s
will also be a holonomy element of the form
\begin{equation}
\label{D}
U_1 U_2 U_3 \cdots U_{n-1} U_n= e^{\frac i2 \not{N}},
\end{equation}
where
\begin{equation}
\label{Db}
\not\!\!{N} \equiv {\bf N \cdot \sigma} \equiv N^i \sigma_i
 = \left( \begin{array}{lr}
 N^3        & N^1 + iN^2 \\
 N^1 - iN^2 &      - N^3
 \end{array} \right)
\end{equation}
for some vector ${\bf N}$
with Cartesian coordinates $N^i$ (and which without loss
of generality can be taken to have length, which represents
the total angle of rotation, $\leq 2\pi$).
Then the constraint (\ref{B}) is the condition that the
total rotation vector is zero,
\begin{equation}
\label{Dc}
{\bf N} = 0, \ \mathrm{or} \ N^i = 0 \ \mathrm{for \ each} \ i,
\end{equation}
which gives the three conditions on the deficit angles.

	The constraint Eq. (\ref{B}) has two immediate consequences:
(1) there cannot exists a thorny sphere
with a single conical singularity
with a deficit angle $0<|\mu|<2\pi$, and
(2) on a sphere with a pair of
conical singularities, the singular points lie on the same axis.

	We can note that Eq. (\ref{B}) in the $SU(2)$
spinor representation
follows from the single equation
\begin{equation}\label{4.8}
\mbox{Tr}\left[ U_1 U_2 U_3 \cdots U_{n-1} U_n \right]=2.
\end{equation}
Indeed, the product of matrices in left hand side of Eq. (\ref{D})
is a unitary matrix which corresponds
to a rotation by an angle $\varphi$ around some axis. 
If the trace of this 
matrix is 2, then the angle is $\varphi=4\pi m$,
where $m$ is an integer, and the unitary matrix is simply the unit
matrix.  Conversely, when the trace of a $2\times 2$ unitary
matrix is 2, it must be the unit matrix.

\subsection{Constraint equations for small angle deficits}

	In the case of small deficit angles,
so that all of the sums of the products of different matrices
${\not\!\! n_k}$ are small, then the constraint Eq. (\ref{B})
or (\ref{Dc} becomes
\begin{equation}
\label{D2}
 0 = {\bf N} \approx \sum_{k=1}^n \mu_k {\bf n}_k,
\end{equation}
which is Eq. (\ref{i1}) of the Introduction.
Since in the three-dimensional space with small deficit angles,
the conical singularities
correspond to strings with tension $\mu_k$ at directions
given by $\bf{n}_k$, the constraint equation becomes the
equilibrium condition for the forces exerted by the strings,
a force-balance equation. For an even number of conical 
singularities (\ref{D2})
is satisfied when for each conical singularity
whose position is defined by the vector ${\bf n}_k$,
there is another singularity with the same deficit angle
whose position is  $-{\bf n}_k$. This situation is realized
for polyhedral configurations of singularities discussed in 
\cite{FrFu:01}.

\section{Construction of thorny spheres with large deficits: Examples}
\setcounter{equation}0

\subsection{Three conical singularities}

\subsubsection{Case of equal angle deficits}

We now discuss some examples of spheres with large deficit
angles.
To construct them it is enough
to solve the constraint equation (\ref{B}).
Let us consider first the case
in which there are three deficit angles that are all the same
and equal to $\mu$,
and the conical singularities are
at points ${\bf n}_1$, ${\bf n}_2$, ${\bf n}_3$.
The constraint Eq. (\ref{B})
for this configuration yields the holonomy around one point in terms of 
holonomies of two other points.  We will write this 
in the following form, using the $SU(2)$ representation
of the holonomy given by Eq. (\ref{C}):
\begin{equation}\label{L2}
U({\bf n}_1,\mu)~
U({\bf n}_2,\mu)=U^{-1}({\bf n}_3,\mu)=U(-{\bf n}_3,\mu).
\end{equation}
Suppose that the angle between vectors ${\bf n}_2$ and ${\bf n}_1$
is $a$ ($0< a <\pi$). We can choose ${\bf n}_1$ and 
${\bf n}_2$ lying in the
(xy)-plane such that matrices for the corresponding points are
\begin{equation}\label{L3}
{\not\!\! n}_1=\cos a ~\sigma_1+\sin a ~\sigma_2,~~
{\not\!\! n}_2=\sigma_1.
\end{equation}
We get 
\begin{eqnarray}\label{L4}
U({\bf n}_1,\mu)~
U({\bf n}_2,\mu)=\left(\cos^2{\mu \over 2}-\cos a ~
\sin^2{\mu \over 2}\right)~I \nonumber \\
+i\sin{\mu \over 2}\left[\cos{\mu \over 2}(1+\cos a)~\sigma_1
+\cos{\mu \over 2}\sin a ~\sigma_2+\sin{\mu \over 2}\sin a ~ 
\sigma_3\right].
\end{eqnarray}
To satisfy (\ref{L2}) one  has  to choose
$a$ such that
\begin{equation}\label{L5}
\cos a =-{\cos{\mu \over 2} \over 1 +\cos{\mu \over 2}}.
\end{equation}
In the limit that $\mu = 0$, one gets $a = 4\pi/3$,
so each spherical triangle fills a hemisphere.
As $\mu$ is increased, the length $a$ of the sides decreases
and reaches 0 when $\mu = 4\pi/3$.

\begin{figure}
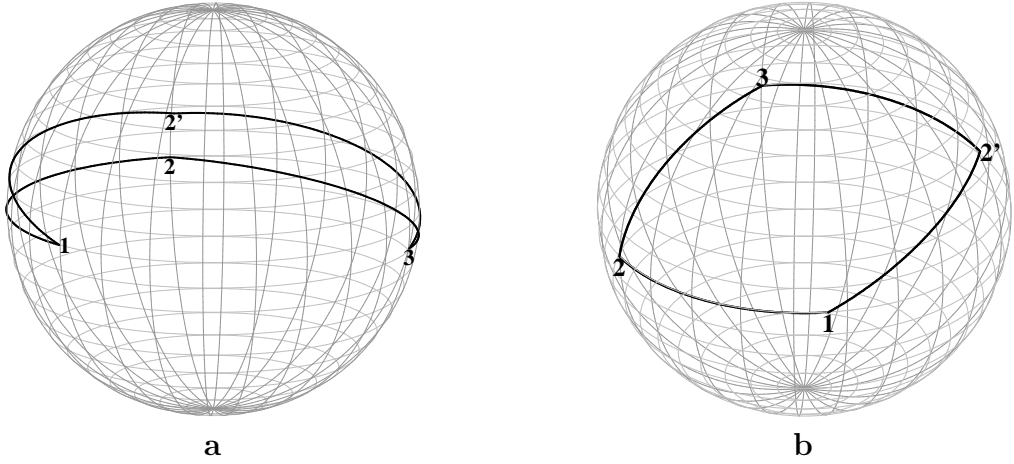

\begin{center}
\[
\begin{array}{cc}
\hspace{-1cm}
\epsfig{file=3pa01.eps, height=5.5cm}
&
\hspace{2cm}\epsfig{file=3pb01.eps, height=5.5cm}\\
\hspace{-1cm}{\bf a} & \hspace{2cm}{\bf b}
\end{array}
\]
\end{center}
\caption[f1]{Figures (a) and (b) demonstrate cuts
which after gluing yield thorny spheres with
three conical singularities with deficits $\mu=\pi/12$
and $7\pi/6$, respectively. On figure (a) points 1 and 3 lie
on the back side of the sphere. 
}
\label{large-def}
\end{figure}

Given (\ref{L5}), the position of the third point, ${\bf n}_3$, is fixed
by the matrix which follows from (\ref{L2}), 
\begin{equation}\label{L6}
{\not\!\! n}_3=\cos a ~ \sigma_1-\cos{\mu \over 2}\sin a ~\sigma_2-
\sin{\mu \over 2}\sin a ~\sigma_3.
\end{equation}
Eqs. (\ref{L3}) and (\ref{L6}) give coordinates of the conical
singularities with deficit angle $\mu$. The corresponding
sphere with conical singularities can be constructed
from a spherical polygon
${\cal M}_C$ with a boundary $C$ on regular sphere $S^2$.
This can be done by either method A or B.  
Consider, for instance, method A. The contour $C$
consists of two parts, $C_1$ and $C_2$.  The contour $C_1$
consists of two shortest geodesics connecting
points ${\bf n}_1$ with ${\bf n}_2$ and ${\bf n}_2$ with ${\bf n}_3$.
The contour $C_2$ consists of geodesics connecting
${\bf n}_1$ with ${\bf n}'_2$ and ${\bf n}'_2$ with ${\bf n}_3$
where ${\bf n}'_2$ is the image of ${\bf n}_2$ 
obtained by rotation around ${\bf n}_1$ by angle $\mu$.
\begin{equation}\label{L7}
{\not\!\! n}_2{\ }'
=U({\bf n}_1,\mu)~{\not\!\! n}_2~U^{-1}({\bf n}_1,\mu).
\end{equation}
For some values of $\mu$ and $a$ the corresponding 
contours are presented in Fig. \ref{large-def}.

\subsubsection{Arbitrary angle deficits $\mu_k$}

	In the general case of $n=3$ conical singularities
with arbitrary deficit angles $\mu_k$,
let ${\bf A} = {\bf n}_1\tan{\frac{\mu_1}2} = \pm A \, {\bf n}_1$,
${\bf B} = {\bf n}_2\tan{\frac{\mu_2}2} = \pm B \, {\bf n}_2$, and
${\bf C} = {\bf n}_3\tan{\frac{\mu_3}2} = \pm C \, {\bf n}_3$,
where $A$, $B$, and $C$ are chosen to be positive.
Then
\begin{equation}
\label{E0}
 U_1 = U({\bf n}_1,\mu_1) = \cos{\frac{\mu_1}2}(I+i{\not\!\! A})
  = {I+i{\not\!\! A}\over\sqrt{1+A^2}},
\end{equation}
\begin{equation}
\label{E2}
 U_2 = U({\bf n}_2,\mu_2) = \cos{\frac{\mu_2}2}(I+i{\not\!\! B})
  = {I+i{\not\!\! B}\over\sqrt{1+B^2}},
\end{equation}
\begin{equation}
\label{E3}
 U_3 = U({\bf n}_3,\mu_3) = \cos{\frac{\mu_3}2}(I+i{\not\!\! C})
  = {I+i{\not\!\! C}\over\sqrt{1+C^2}},
\end{equation}
where the signs of the square roots in the denominators
are chosen to be the same as those of the respective
$\cos{\frac{\mu_k}2}$'s (positive if $|\mu_k| < \pi$).
Then for $U_1 U_2 U_3$ to be the unit matrix that represents
the trivial holonomy, one uses the Pauli-matrix identity
(for $i\sigma_1\sigma_2\sigma_3=1$)
\begin{equation}
\label{E4}
 {\not\!\! A}{\not\!\! B}
  \equiv ({\bf A \cdot \sigma})({\bf B \cdot \sigma})
 = ({\bf A}\!\cdot\!{\bf B})\,I - i({\bf A \times B )\cdot \sigma}
\end{equation}
and gets the constraint
\begin{equation}
\label{F}
 \bf{A}+\bf{B}+\bf{C}+\bf{A}\times\bf{B}+\bf{A}\times\bf{C}
 +\bf{B}\times\bf{C}-(\bf{A}\cdot\bf{B})\,\bf{C}
 +(\bf{A}\cdot\bf{C})\,\bf{B}-(\bf{B}\cdot\bf{C})\,\bf{A} = 0.
\end{equation}
One can see that the linear part of this is simply that
the sum of the three vectors is zero, $\bf{A}+\bf{B}+\bf{C} = 0$.

	In the nonlinear case, if, say, $\bf{A}$ and $\bf{B}$
are specified, then the solution for $\bf{C}$ is
\begin{equation}
\label{G}
 \bf{C} = - \frac{\bf{A}+\bf{B}+\bf{A}\times\bf{B}}
 {1-\bf{A}\cdot\bf{B}}.
\end{equation}
Again one can see that if $\bf{A}$ and $\bf{B}$ are small,
to linear order in those vectors, $\bf{C} = - \bf{A} - \bf{B}$.
Alternatively, if the three unit normals $\bf{n}_k$ are
specified, then the solution for the deficit angles is
\begin{equation}
\label{H}
 \tan{\frac{\mu_1}2} = \frac{\bf{n}_1\cdot(\bf{n}_2\times\bf{n}_3)}
 {(\bf{n}_1\times\bf{n}_2)\cdot(\bf{n}_1\times\bf{n}_3)},
\end{equation}
\begin{equation}
\label{H2}
 \tan{\frac{\mu_2}2} = \frac{\bf{n}_1\cdot(\bf{n}_2\times\bf{n}_3)}
 {(\bf{n}_2\times\bf{n}_1)\cdot(\bf{n}_2\times\bf{n}_3)},
\end{equation}
\begin{equation}
\label{H3}
 \tan{\frac{\mu_3}2} = \frac{\bf{n}_1\cdot(\bf{n}_2\times\bf{n}_3)}
 {(\bf{n}_3\times\bf{n}_1)\cdot(\bf{n}_3\times\bf{n}_2)}.
\end{equation}

	A third specification would be to fix the three
deficit angles and thereby to fix the three lengths
$A$, $B$, and $C$ of the three vectors $\bf{A}$, $\bf{B}$, and $\bf{C}$
respectively.  Then the constraint determines the relative
directions of $\bf{A}$, $\bf{B}$, and $\bf{C}$.
Suppose that these are given by the cosines of the angles
between them, say
\begin{equation}
\label{I}
 \alpha = \cos{a} = {\bf n_2 \!\cdot\! n_3} = {\bf B \!\cdot\! C}/(BC),
\end{equation}
\begin{equation}
\label{J}
 \beta = \cos{b} = {\bf n_3 \!\cdot\! n_1} = {\bf C \!\cdot\! A}/(CA),
\end{equation}
\begin{equation}
\label{K}
 \gamma = \cos{c} = {\bf n_1 \!\cdot\! n_2} = {\bf A \!\cdot\! B}/(AB).
\end{equation}
Then by equating the squared magnitudes of the two sides of
Eq. (\ref{G}), one can solve for
\begin{equation}
\label{L}
 \gamma={1\over AB}\left[1\mp\sqrt{1+A^2+B^2+A^2B^2\over 1+C^2}\right].
\end{equation}
Similarly, by cyclic permutations one gets
\begin{equation}
\label{M}
 \alpha={1\over BC}\left[1\mp\sqrt{1+B^2+C^2+B^2C^2\over 1+A^2}\right],
\end{equation}
\begin{equation}
\label{N}
 \beta={1\over CA}\left[1\mp\sqrt{1+C^2+A^2+C^2A^2\over 1+B^2}\right].
\end{equation}

	One must choose the signs so that the cosines are between
$-1$ and 1.  (For small $A$, $B$, and $C$,
one must choose the upper signs.)
In order that the cosines can be between $-1$ and 1,
the absolute magnitudes of the three deficit angles
must obey the triangular inequalities
(each larger than the absolute difference between the other two,
and each smaller than the sum of the other two),
which translates into the nonlinear inequalities for $A$, $B$, and $C$
that
\begin{equation}
\label{O}
 {|A-B|\over |1+AB|} \leq C \leq {A+B \over |1-AB|},
\end{equation}
\begin{equation}
\label{P}
 {|B-C|\over |1+BC|} \leq A \leq {B+C \over |1-BC|},
\end{equation}
\begin{equation}
\label{Q}
 {|C-A|\over |1+CA|} \leq B \leq {C+A \over |1-CA|}.
\end{equation}

	In the case of three conical singularities
that we are presently considering, the regions ${\cal M}_1$
and ${\cal M}_2$ are simply spherical triangles that
are identical except for their orientation.
The interior angles at the vertices $\bf{A}$, $\bf{B}$, and $\bf{C}$
are $\pi-\mu_1/2$, $\pi-\mu_2/2$, and $\pi-\mu_3/2$ respectively,
and the cosines of the angular lengths of the opposite sides
are $\alpha = \cos{a}$, $\beta = \cos{b}$, and $\gamma = \cos{c}$
respectively.
Then, as an alternative to the constraint equations given above,
one can use the standard formulas (\ref{0.3a}) and (\ref{0.3})
for spherical triangles.
For example, from Eq. (\ref{0.3a}),
\begin{equation}
\label{V}
 \alpha = \cos{a}
  = {\cos{\mu_2 \over 2}\cos{\mu_3 \over 2} - \cos{\mu_1 \over 2}
  \over \sin{\mu_2 \over 2}\sin{\mu_3 \over 2}}
\end{equation}
and cyclically for $\beta = \cos{b}$ and $\gamma = \cos{c}$.
This reduces to Eq. (\ref{L5}) in the special case
in which all of the three deficit angles are equal to $\mu$.

\subsection{Four conical singularities}

\subsubsection{Case of equal angle deficits}

Consider now a sphere with four conical singularities with 
deficits $\mu_k$ at points ${\bf n}_k$, $k=1,2,3,4$.
The constraint (\ref{B}) on the holonomies is 
\begin{equation}\label{L8}
U({\bf n}_1,\mu_1)~U({\bf n}_2,\mu_2)~U({\bf n}_3,\mu_3)~
U({\bf n}_4,\mu_4)=I.
\end{equation}

\begin{figure}
\centerline{\epsfig{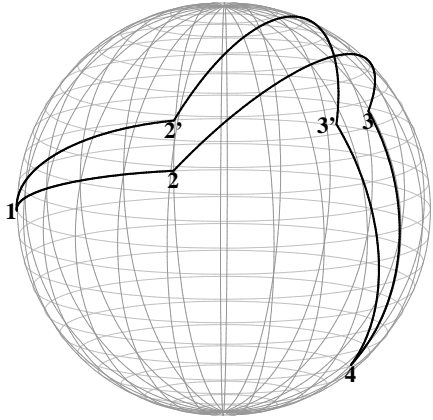}}
\caption[f2]{This figure shows the cut for 4 conical
singularities with deficits $\mu=\pi/12$ and parameters
$\cos a=0.2$ and $\theta=\pi/2$. Points 3 and 3' lie on the back side
of the sphere.}
\label{large4-def}
\end{figure}

Given the coordinates and angular deficits of three points, we can find
from (\ref{L8}) the coordinates and the deficit of the
fourth point. We first present here a particular solution
of (\ref{L8}) for the case in which all the deficit angles coincide.  
Then the constraint (\ref{L8}) can be rewritten as
\begin{equation}\label{L9}
U({\bf n},\beta)=U({\bf n}_1,\mu)~U({\bf n}_2,\mu)=U(-{\bf n}_4,\mu)~
U(-{\bf n}_3,\mu)~.
\end{equation}
where ${\bf n}$ is a unit vector in the Euclidean 3-space.
The parameters $\beta$ and $\bf n$
are uniquely defined by (\ref{L9}) if $\mu$ and the coordinates
of the pair of points ${\bf n}_1$, ${\bf n}_2$ are known.
For ${\bf n}_1$, ${\bf n}_2$ defined as in (\ref{L3}),
\begin{equation}\label{L11}
{\not\!\! n}=C^{-1/2}\left(\cos{\mu \over 2} (1+\cos a)~\sigma_1+
\cos{\mu \over 2}\sin a~\sigma_2-\sin{\mu \over 2} \sin a
~\sigma_3\right),
\end{equation}
\begin{equation}\label{L11a}
C=\sin^2 a+\cos^2{\mu \over 2}(1+\cos a)^2.
\end{equation}
Given coordinates ${\bf n}_1$ and ${\bf n}_2$ 
it is easy to see that (\ref{L9}) holds if
${\bf n}_3$ and ${\bf n}_4$ are obtained
by a rotation of $-{\bf n}_2$, $-{\bf n}_1$ around 
$\bf n$ by some angle $\theta$, i.e.,
\begin{equation}\label{L12}
{\not\!\! n}_3=-U({\bf n},\theta)~{\not\!\! n}_2~U^{-1}({\bf n},\theta),
~~~
{\not\!\! n}_4=-U({\bf n},\theta)~{\not\!\! n}_1~U^{-1}({\bf n},\theta).
\end{equation}
This procedure yields a three-parameter family of spheres
with four conical singularities. The parameters are $\mu$, $\theta$
and the angle $a$ between ${\bf n}_1$ and ${\bf n}_2$.
The internal region ${\cal M}_C$ of an example of such a sphere
obtained by method B is 
shown in Fig. \ref{large4-def}.
Points
${\bf n}_2'$ and ${\bf n}_3'$ are the images of ${\bf n}_2$
and ${\bf n}_3$,
respectively. The same method was applied to get
configurations with 6 conical singularities on Figures \ref{A2}
and \ref{MB}.

\subsubsection{Arbitrary angle deficits $\mu_k$}

	Now let us turn to the case when the $n=4$ conical singularities
have arbitrary conical deficit angles $\mu_k$.
Similar to what was done for $n=3$,
let ${\bf A} = {\bf n_1}\tan{\frac{\mu_1}2} = \pm A {\bf n_1}$,
${\bf B} = {\bf n_2}\tan{\frac{\mu_2}2} = \pm B {\bf n_2}$,
${\bf D} = {\bf n_3}\tan{\frac{\mu_3}2} = \pm D {\bf n_3}$,
${\bf E} = {\bf n_4}\tan{\frac{\mu_4}2} = \pm E {\bf n_4}$,
where $A$, $B$, $D$, and $E$ are chosen to be positive.
Now there are various ways to proceed with solving the constraint
$U_1 U_2 U_3 U_4 = I$, depending on what is specified and
what is to be solved for.

	If $\bf{A}$, $\bf{B}$, and $\bf{D}$ are specified
and $\bf{E}$ is to be solved for, one writes the constraint
in the form
\begin{equation}
\label{E1}
 U_4 \equiv {I+i{\not\!\! E}\over\sqrt{1+E^2}}
 = U_3^{-1} U_2^{-1} U_1^{-1}
 = {(I-i{\not\!\! D})(I-i{\not\!\! B})(I-i{\not\!\! A}) \over
    \sqrt{1+D^2}\sqrt{1+B^2}\sqrt{1+A^2}},
\end{equation}
and then one can explicitly solve for
\begin{equation}
\label{F1}
 \bf{E} = {- \bf{A} - \bf{B} - \bf{D}
          - \bf{A}\times\bf{B}-\bf{A}\times\bf{D}-\bf{B}\times\bf{D}
	  +(\bf{A}\cdot\bf{B})\,\bf{D}-(\bf{A}\cdot\bf{D})\,\bf{B}
	  +(\bf{B}\cdot\bf{D})\,\bf{A}
    \over 1-\bf{A}\cdot\bf{B}-\bf{A}\cdot\bf{D}-\bf{B}\cdot\bf{D}
           -(\bf{A}\times\bf{B})\cdot\bf{D}}.
\end{equation}

	This obviously generalizes to arbitrary $n$:
If the first $n-1$ positions and deficit angles are specified,
so that $U_k$ is given for $1 \leq i \leq n-1$, then
one can readily solve for
\begin{equation}
\label{G1}
 U_n = U_{n-1}^{-1} U_{n-2}^{-1} \cdots U_2^{-1} U_1^{-1}.
\end{equation}

	If instead all of the deficit angles are given,
and all but two consecutive positions on the sphere, then,
assuming that the appropriate triangular inequalities are satisfied,
one can solve for these two positions up to an overall
rotation about an axis determined by what is given.
(This rotation has nontrivial significance only for $n \geq 4$.)
Let us illustrate this with the case $n=4$.

	Suppose that the positions and deficit angles of the
3rd and 4th conical singularities are given,
so that $\bf{D}$ and $\bf{E}$ are given,
and the deficit angles but not the positions of the
1st and 2nd conical singularities are given,
so that $A$ and $B$ are given, but not the direction
$\bf{n_1}$ of $\bf{A}$ and the direction $\bf{n_2}$ of $\bf{B}$.
Then we can use the constraint in the form
\begin{equation}
\label{H1}
 U_2^{-1} U_1^{-1} = U_3 U_4,
\end{equation}
or
\begin{equation}
\label{I1}
 {I - {\bf A}\cdot{\bf B}
    - i({\bf A}+{\bf B}+{\bf A}\times{\bf B})\cdot{\bf \sigma} \over
  \sqrt{1+A^2}\sqrt{1+B^2}}
  = U_3 U_4 = {I + i{\not\!\! C} \over \sqrt{1+C^2}},
\end{equation}
where for $n=4$
\begin{equation}
\label{J1}
 {\bf C} = {{\bf D + E + D\times E}\over 1-{\bf D\cdot E}}
\end{equation}
represents the combined effect of $U_3$ and $U_4$.

	Then we can proceed as we did for $n=3$
when the magnitudes $A$, $B$, and $C$ were given,
to solve for the angles between $\bf{A}$, $\bf{B}$, and $\bf{C}$.
Since here $\bf{C}$ is determined from the given $\bf{D}$ and $\bf{E}$,
the directions of $\bf{A}$ and $\bf{B}$ are then determined
up to an overall rotation about the vector $\bf{C}$.

\subsection{$n$ conical singularities}

	Obviously, the procedure of the preceding section
generalizes to higher $n$.
If the $U_k$ are specified for all $3 \leq i \leq n$,
and if the deficit angles $\mu_1$ and $\mu_2$
are also specified (so that
$A = |\tan{\mu_1/2}|$ and $B = |\tan{\mu_2/2}|$ are specified),
then we define $\bf{C}$ by
\begin{equation}
\label{K1}
  U_3 U_4 \cdots U_{n-1} U_n = {I + i{\not\!\! C} \over \sqrt{1+C^2}}
\end{equation}
and solve for the angles between $\bf{A}$, $\bf{B}$, and $\bf{C}$.
This determines $\bf{A}$ and $\bf{B}$
up to an overall rotation about the vector $\bf{C}$.

	If all of the deficit angles are specified,
and all but two non-consecutive positions,
then one has to permute the positions and put
in the appropriate commutators to solve for those positions
(up to the arbitrary rotation).
For example, for $n=4$, suppose that the deficit angles but not
the positions of the 1st and 3rd points are specified,
and that both are specified for the 2nd and 4th points,
so that $U_2$ and $U_4$ are given.
Then if we define $\tilde{U}_3 = U_2 U_3 U_2^{-1}$,
the constraint equation becomes
$\tilde{U}_3^{-1} U_1^{-1} = U_2 U_4$,
and one can use the previous procedure to solve
for $U_1$ and $\tilde{U}_3$, up to an overall rotation
about the axis of the rotation $U_2 U_4$.
Finally, one reconstructs $U_3 = U_2^{-1} \tilde{U}_3 U_2$.
This procedure has a straightforward generalization for all higher $n$
and for the unspecified positions to be further separated
in the cyclic chain of rotations $U_k$
that combine to produce the identity.

\section{Topological aspects of sphere cutting}  
\setcounter{equation}0

Using method $A$ developed in Section 3 one gets a map 
\begin{equation}\label{6.2}
\Psi: \tilde{S}^2 \to \{{\cal M}_1, {\cal M}_2\}\, 
\end{equation} 
of a thorny sphere $\tilde{S}^2$ onto a pair of simply connected
regions on a unit round sphere. The boundaries ${\cal B}_1=\partial 
{\cal M}_1$ and ${\cal B}_2=\partial  {\cal M}_2$ are isometric
spherical polygons which are to be identified. Each of the polygons
has  $n$ vertices, $A_{(1,k)}$ and $A_{(2,k)}$, where $n$ is a
number of the conical singularities of the thorny sphere $\tilde{S}^2$.
It should be emphasized that the change of the reference point which is
used to order the conical singularities may result in the change of
order of vertices $A_{(1,k)}$ and $A_{(2,k)}$, and as a result of
this, one can get different choice of regions ${\cal M}_1, {\cal M}_2$
representing the same thorny sphere. It is evident that the
corresponding maps $\Psi$ and $\Psi'$ formally being different,
are in fact equivalent.

When we construct regions  ${\cal M}_1$ and ${\cal M}_2$ using the
method described in section 4.1 by solving the constraint equations we
do not know in advance which ordering procedure of the conical
singularities on the thorny sphere would correspond to the set of
vertices obtained by gluing the boundaries ${\cal B}_1$ and ${\cal
B}_2$ and identifying the vertices $A_{(1,k)}$ and $A_{(2,k)}$. In
fact, the situation is even more complicated. 

Let us note that the constraint equations (\ref{n10}) and (\ref{B})
remain unchanged if one includes a unit operator between any two
subsequent terms, say $i$ and $i+1$ in the product of matrices. But a
rotation along an axis ${\bf n}$ by the angle $2\pi m$ is represented
by a unit operator. Thus adding two new vertices with angle deficits
$2\pi m$ (one for each of the regions ${\cal M}_k$) does not violate the
constraint equations. 
We can choose the new angle deficits to be
negative and put the new vertices at the north and south poles of a
round sphere. This is equivalent to the usage of a covering space
$S^2_m$ for a round sphere with a winding number $m>0$. That is why by
solving the constraint equations (\ref{n10}) and (\ref{B}) one may end
not with regular regions ${\cal M}_1, {\cal M}_2$ on a round sphere,
but with regions on a  covering space for $S^2_m$. After identifying
the points of $m$ different leaves and projecting  $S^2_m$ onto $S^2$,
one obtains boundaries ${\cal B}_1$ and ${\cal B}_2$ which are
topologically circles $S^1$, but which have intersections.

In order to exclude such cases one must be certain that after a
solution of the constraint equations one does not have undesirable
extra vertices with $-2\pi m$ angle deficits.
We describe now a procedure which allows one to do this.  

Consider a map $\psi_a$ of the region ${\cal M}_a$ onto the region
$\tilde{{\cal M}}_a$ of the thorny sphere
\begin{equation}\label{6.3}
\psi_a: {\cal M}_a \to \tilde{{\cal M}}_a \, . 
\end{equation} 
Under this map the boundary ${\cal B}_a$ of ${\cal
M}_a$ is transformed into the boundary $\tilde{{\cal B}}_a$
of $\tilde{{\cal
M}}_a$. Take a  regular point $\tilde{p}$
inside $\tilde{{\cal M}}_a $ and let $p_k$ be its images under
$\psi_a^{-1}$. Let $y^{\mu}$ be coordinates near $\tilde{p}$
and $x^{\nu}_k$ be
coordinates near ${p}_k$. Since $\tilde{{\cal M}}_a$ and 
${\cal M}_a $ are orientable manifolds, we use local coordinate systems
on both
of them so that any transition from one coordinate system to another on
the same manifold has the value of the Jacobian equal to $+1$.
The {\em
degree of a map} $\psi_a$ is determined as
\begin{equation}\label{6.4}
\mbox{deg}\, \psi_a = \sum_{k}\, \mbox{sign}\det \left( {\partial
y^{\mu}\over \partial x_k^{\nu}}  \right)_{{p}_k}\, . 
\end{equation}

One can prove (see e.g. \cite{DuFoNo:85,Fran}) that the index of the map
does not
depend on the choice of the regular point $\tilde{p}$ and is invariant
under smooth homotopies. Moreover, the index of the map $\psi_a$ is the
same as the index of the map $\psi_a$ restricted to the boundary ${\cal
B}_a$. Since both of the boundaries are topologically circles $S^1$,
the degree of this map is just a winding number $m$.
Note that ${\cal M}_1$ and ${\cal M}_2$ have the same degree of map
because ${\cal B}_2$  can be obtained by a deformation 
of ${\cal B}_1$ (recall that we got vertices on ${\cal B}_2$
by rotating vertices on ${\cal B}_1$).

To calculate this winding number we consider a stereographic projection
of ${\cal B}_a$ into a plane. We take the origin of the stereographic projection to lie in the interior of the region ${\cal M}_1$. For a resulting curve ${\hat{\cal B}}_a$ 
with intersections we define 
\begin{equation}\label{6.5}
m = {1\over 2\pi}\, \oint \, k \, dl\, , 
\end{equation}
where $k$ is the Gaussian curvature of the curve ${\hat{\cal B}}_a$ and
$dl$ is the proper distance element. Since  $m$ is invariant under
smooth homotopic transformations, it remains the same under a
continuous change of the position of the `north' pole used for the
stereographic projection until it cross ${\cal B}_a$.

To summarize the above discussion we stress that using method $A$,
starting with a given thorny sphere $\tilde{S}^2$ one can define
(not uniquely)
regions ${\cal M}_{1,2}$ and the gluing procedure which recovers
$\tilde{S}^2$. In the inverse procedure, when one starts with a 
solution of the constraint equations, one must first check
whether the winding number of the boundary is 1.
Only in this case will the gluing procedure
give a thorny sphere $\tilde{S}^2$ without any additional
angle deficits that are negative integer multiples of $2\pi$.

\section{Thorny spheres and solutions of Einstein equations
with radial strings}
\setcounter{equation}0

Let us discuss now solutions of the Einstein equations for the
thorny sphere configurations,
with strings at the conical singularities.
The total action of the system 
is\footnote{In this section we restore a
normal value $G$ for the Newton constant.}
\begin{equation}\label{1.9}
I={1 \over 16\pi G}\left[\int_{\cal M} \sqrt{-g}d^4x R+
2\int_{\partial {\cal M}}K\sqrt{-h}d^3x\right]-
{1 \over 4\pi}\sum_k \hat{\mu}_k\int\sqrt{\sigma_k}\,d^2\zeta_k~~~.
\end{equation}
The last term in the right hand side of (\ref{1.9})  is the Nambu-Goto
action for the strings, where $(\sigma_k)_{\alpha\beta}$ 
is the metric induced on the
world-sheet of a particular string. We assume in (\ref{1.9}) that the
space-time $\cal M$ has a  time-like boundary.  We take the metric in
the form
\begin{equation}\label{1.10}
ds^2=\gamma_{\alpha \beta}dx^\alpha dx^\beta+e^{2\phi}
a^2\, d{\tilde\Omega}^2~~~.
\end{equation}
Here $\gamma_{\alpha\beta}$ is a 2D metric, $\phi=\phi(x)$ a dilaton
field which depends on coordinates $x^\alpha$,
and $d{\tilde\Omega}^2$ is the
metric on the thorny sphere $\tilde{S}^2$ with conical singularities.
For a string located at fixed angles,
the induced metric on  a string world-sheet coincides with 
$\gamma_{\alpha\beta}$. The parameter $a$ in (\ref{1.10}) has the
dimensionality  of the length. Locally near each string the metric
$d{\tilde\Omega}^2$ can be written as
\begin{equation}\label{1.11}
d{\tilde\Omega}^2=\sin^2\theta d\varphi^2+d\theta^2~~~,
\end{equation}
where $0\leq \theta \leq \pi$, and $\varphi$ is periodic
with period $2\pi-\mu_k)$.
To proceed we have to take into account in (\ref{1.9}) 
the presence of 
delta-function-like contributions due to the conical 
singularities (see, for instance \cite{FS})
\begin{equation}\label{1.13}
\int_{\cal M} \sqrt{-g}d^4x R=
\int_{{\mbox{\small{reg}}} (M^2)} \sqrt{-g}d^4x R
+2\sum_k\mu_k \int 
\sqrt{\sigma_k}\,d^2\zeta_k~~,
\end{equation}
where reg$(M^2)$ is the regular domain of $\cal M$.
If we impose the on-shell condition $\mu_k=8\pi\, G\hat{\mu}_k$,
the contribution of the conical singularities in the
curvature in (\ref{1.9}) will cancel exactly the 
contribution from the string actions.
There will remain only the bulk
part of the action.
On the metric (\ref{1.10}) it will reduce to
the 2D dilaton gravity action
\begin{equation}\label{1.14}
I={1 \over 4G_2} \left[\int \sqrt{\gamma}d^2x\left(
e^{2\phi}R_2+2e^{2\phi}(\nabla\phi)^2+{2 \over a^2}\right)
+2\int dy e^{2\varphi}(k-k_0)\right]~~~,
\end{equation}
\begin{equation}\label{1.15}
{1 \over G_2}={a^2 C\over G}
\end{equation}
\begin{equation}\label{1.15b}
C = 1-2G\sum_k \hat{\mu}_k = 1 - {1 \over 4\pi}\sum_k \mu_k~~~.
\end{equation}
The curvature $R_2$ in (\ref{1.14}) is the 2D curvature
determined by $\gamma_{\alpha\beta}$.
As a result of the modification of the area of sphere due to
the conical singularities,
the gravitational action (including the boundary term)
acquires an overall coefficient 
which depends on the $\mu_k$'s.
We included this coefficient in the
definition of effective two dimensional gravitational 
coupling $G_2$, Eq. (\ref{1.15}).
It is important that the action (\ref{1.14}) has 
precisely the same form as the
dilatonic action obtained under a spherical reduction of the
gravitational action in the absence of cosmic strings.
Therefore strings have no effect on the dynamical
equations for the metric $\gamma_{\alpha\beta}$ and the
dilaton $\phi$. For these quantities one has 
standard solutions. In particular, the Birkhoff theorem can be 
applied in this case and guarantees that in the absence of 
the other matter in the bulk, the solution is static
and is a 2D black hole of mass $M$.
\begin{equation}\label{bh}
d\gamma^2=-Fdt^2+F^{-1}dr^2~~,~~~F=1-{2M \over r}~~~.
\end{equation}
The corresponding four-dimensional solution is a
Schwarzschild black hole of the same mass parameter,
but with strings in the radial direction.
In a similar way, by using
(\ref{1.14}) one can  construct non-static solutions in the presence of
strings. Non-vacuum static spherically symmetric solutions,
such as a charged black hole with strings, can be constructed as well
by adding matter in the bulk.

For example, if we define $d^2 \hat{\Omega} = d{\tilde\Omega}^2/C$
to give a rescaled thorny sphere with area $4\pi$ and smooth part
having Gaussian curvature no longer unity but
\begin{equation}\label{gc}
K = C = 1-2G\sum_k \hat{\mu}_k = 1 - {1 \over 4\pi}\sum_k \mu_k~~~,
\end{equation}
then the Reissner-Nordstrom black hole generalizes to the following
solution with strings:
\begin{equation}\label{cbh}
ds^2=-\left( K - {2M \over r} + {Q^2 \over r^2} \right) dt^2
     +\left( K - {2M \over r} + {Q^2 \over r^2} \right)^{-1}dr^2
     + r^2 d^2 \hat{\Omega}~~.
\end{equation}
Here $Q$ is precisely the charge, defined as $1/(4\pi)$
times the flux of electric field through each thorny sphere.

This form of the Reissner-Nordstrom metric remains valid
(but is no longer asymptotically flat with a static timelike
Killing vector $\partial/\partial t$)
when the rescaled thorny sphere with positive Gaussian curvature,
$K > 0$, on its smooth part,
is replaced by a rescaled {\em thorny pseudosphere} with 
negative Gaussian curvature, $K < 0$, on its smooth part that is then
locally isometric to a hyperbolic 2-space with constant negative
curvature.

\bigskip

\noindent
\section*{Acknowledgments}

\indent This work was partially supported  by the Natural Sciences and
Engineering Research Council of Canada, the Killam Trust and the NATO Collaborative Linkage Grant CLG.976417.

\newpage

\appendix

\section{Gaussian normal maps and generalized constraint equation}
\setcounter{equation}0

\subsection{Gaussian map}

Let $M^2$ be a closed 2-dimensional surface
in a 3-dimensional Euclidean space
\begin{equation}\label{a.1}
F(X^1,X^2,X^3)=0\, ,\hspace{0.5cm}\nabla F\ne 0\, ,
\end{equation}
or locally $X^{i}=X^{i}(y^1,y^2)$.
Using the rotational freedom in the choice of $X^i$,
one can describe the surface $M^2$ locally as follows:
\begin{equation}\label{a.2}
X^3=f(y^1,y^2)\, ,\hspace{0.5cm}X^1=y^1\, ,\hspace{0.5cm}X^2=y^2\, .
\end{equation}
The first quadratic form (induced metric) is
\begin{equation}\label{a.3}
ds^2=g_{ab}\, dy^a\, dy^b\, ,
\end{equation}
\begin{equation}\label{a.4}
g_{11}\, = 1+f_{,1}^2\, ,\hspace{0.5cm}
g_{22}\, = 1+f_{,2}^2\, ,\hspace{0.5cm}
g_{11}\, = f_{,1}\, f_{,2}\, ,\hspace{0.5cm}
\det g\, = 1+f_{,1}^2 +f_{,2}^2\, ,
\end{equation}
while the components of the second quadratic form are
\begin{equation}\label{a.5}
b_{ab}\, = {f_{,ab}\over \sqrt{\det g}}\, .
\end{equation}
The Gaussian curvature is
\begin{equation}\label{a.6}
K = {\det b\over \det g}= {f_{,11}\, f_{,22}-f_{,12}^2
\over (1+f_{,1}^2 +f_{,2}^2)^2}\, .
\end{equation}
We also have
\begin{equation}\label{a.7}
R=2K\, ,
\end{equation}
where $R$ is the Ricci scalar for the induced metric $g_{ab}$.

Consider a unit sphere $S^2$ determined by the equation
\begin{equation}\label{a.8}
(X^1)^2+(X^2)^2+(X^3)^2=1\, .
\end{equation}
Let $(\zeta^1, \zeta^2)$ be  coordinates on the sphere, so 
\begin{equation}\label{a.9}
X^i=X^i(\zeta^a)\, .
\end{equation}
Then the induced metric on the unit $S^2$ in these coordinates is 
\begin{equation}\label{a.10}
\gamma_{ab}=\delta_{ij}\, X^i_{,\zeta^a} \, X^j_{,\zeta^b}\, .
\end{equation}
The area element is 
$da=\sqrt{\det(\gamma)}\, d\zeta^1\,d\zeta^2\,$.

We determine the {\em Gaussian normal map} $\varphi: M^2 \to S^2$
of $M^2$ into the sphere $S^2$ as follows (see e.g. \cite{DuFoNo:85,Fran}).
Let $n^{i}(P)$ be a unit normal to 
$M^2$ at a point $P$. Then we put into correspondence with $P$,
a point $p$  on the unit sphere with coordinates $X^i=n^{i}(P)$
(which means that normal vector to $M^2$ at point $P$ coincides
with the normal vector to $S^2$ at point $p$). 
This map determines the relation between coordinates $y^a$ on $M^2$
and  coordinates $\zeta^a$ on $S^2$
\begin{equation}\label{a.11}
y^a=y^a(\zeta^b)\, .
\end{equation}

Let us now show that for the Gaussian normal map
the following relation is valid \cite{DuFoNo:85,Fran}:
\begin{equation}\label{a.12}
K\,dA  =  da ,
\end{equation}
where $dA=\sqrt{g}\, dy^1\, dy^2$ is the surface area element on $M^2$
and $da$ is the area element on $S^2$.
To prove this relation we choose $X^3$ to be orthogonal to $M^2$
at a given point $p$ and $y^1=X^1$ and $y^2=X^2$ to be tangent
to this surface. Then the surface $M^2$ is determined by the equation
$X^3=f(y^1,y^2)$, where $f_{,1}=f_{,2}=0$ at the point $p$. Hence
\begin{equation}\label{a.13}
K=\det \left| 
\begin{array}{cc}
f_{,11} & f_{,12} \\
f_{,21} & f_{,22} 
\end{array} 
\right|\, ,\hspace{0.5cm}
g_{ab}=\delta_{ab}\, .
\end{equation}
Euclidean coordinates of a unit normal vector $N^i$
in the vicinity of the point $p$ are
\begin{equation}\label{a.14}
n^{i}(y^1,y^2)= (1+f_{,1}^2+f_{,2}^2)^{-1/2}\,
 \left(f_{,1}, f_{,2}, -1\right)\, .
\end{equation}

We choose now the coordinates $\zeta^a$ on $S^2$ so that near a point
$P=\varphi(p)$,
\begin{equation}\label{a.15}
\bar{X}^1=\zeta^1\, ,\hspace{0.5cm}
\bar{X}^2=\zeta^2\, ,\hspace{0.5cm}
\bar{X}^3=\sqrt{1-(\zeta^1)^2-(\zeta^2)^2}\, .
\end{equation}
Here $\bar{X}^i$ are such Cartesian coordinates that axis 
$\bar{X}^3$ coincides with the normal vector to $S^2$ at $p$, $N^i(p)$.
In what follows we will denote normal to $S^2$ the same as
the normal to $M^2$.
In these coordinates $\gamma_{ab}(p)=\delta_{ab}$.
Using (\ref{a.14}) we get
\begin{equation}\label{a.16}
\zeta^a= {f_{,a}\over \sqrt{1+f_{,1}^2+f_{,2}^2}}\, .
\end{equation}
This relation establish a relation between coordinates $y^a$ on $M^2$
and coordinates $\zeta^a$ on $S^2$.
The canonical invariant element of area on a unit sphere at the point
$p$ written in the coordinates $y^a$ is
\begin{equation}\label{a.17}
da= \det \left( \zeta_{,y^b}^a \right)\left|_P \right.\, dy^1\, dy^2
= K\, dy^1\, dy^2=K\sqrt{g}\, dy^1\, dy^2\, ,
\end{equation}
which proves (\ref{a.12}).
To obtain these equalities  we use that $\det g|P=\det \gamma|p=1$.

Suppose that $M^2$ is a compact 2 dimensional manifold diffeomorphic to 
the unit sphere $S^2$ and its embedding in $R^3$ is a closed convex 
surface, so that the Gaussian spherical map is a regular one-to-one map 
on $S^2$. 
In this case one has
\begin{equation}\label{a.18}
\int_{M^2} {\bf n}\, K\, dA =0\, .
\end{equation}
This relation directly follows from (\ref{a.12}) and the relation
\begin{equation}\label{a.19}
\int_{S^2} {\bf n}\, da =0\, .
\end{equation}

\subsection{Generalized constraint equation}

Consider a closed 2D manifold $M^2$  
with $n$ conical
singularities with positive deficit angles $\mu_k$ 
($0<\mu_k <2\pi$), such that $\sum_k\mu_k<4\pi$.
We call such manifold a {\em thorny manifold}, or, for brevity,
a {\em thornifold}.
Assume that the Gaussian curvature $K$ of $M^2$ is positive everywhere,
so that one can isometrically embed the $M^2$ in Euclidean 3-space
as a closed convex  surface.
Consider the Gaussian normal map of a regular domain 
reg$(M^2)$
of $M^2$ onto $S^2$. To see what happens 
under the Gaussian map with conical singularities, 
consider a small region 
$\Sigma_k$ around the $k$th conical singularity (but not
including the singularity itself) with the
boundary ${\cal P}_k$.
The region $\Sigma_k$ is mapped onto a region $\tilde{\Sigma}_k$
on $S^2$ with boundary $\tilde{\cal P}_k$.
When ${\cal P}_k$ shrinks to the conical singularity, the contour
$\tilde{\cal P}_k$ 
shrinks to a contour $C_k$ on $S^2$, because the normal vector
at a conical singularity does not have a unique direction.
Let $D_k$ be the region inside $C_k$. The remarkable 
property of $D_k$ is that, although its form depends
on the concrete thornifold $M^2$, its surface area is $\mu_k$
where $\mu_k$ is the conical angle deficit at the corresponding
singular point. To see this, apply the Gauss-Bonnet formula to
the region $\Sigma_k$
\begin{equation}\label{ba.1}
\int_{\Sigma_k} K dA+\int_{{\cal P}_k} kdl=2\pi~,
\end{equation}
where $k$ is the extrinsic curvature of ${\cal P}_k$
embedded in $M^2$. When ${\cal P}_k$ shrinks to 
the conical singularity, the limit 
for integral $\int_{{\cal P}_k} kdl$ is $2\pi-\mu_k$. So we
get from (\ref{ba.1}) by using  (\ref{a.12})
\begin{equation}\label{ba.2}
A_k=\int_{D_k}da=\lim_{\Sigma_k\rightarrow 0}
\int_{\Sigma_k} K dA
=\mu_k
\end{equation} 
To summarize, the Gaussian map of a thornifold $M^2$ with
$n$ conical singularities is a regular sphere with $n$
disks $D_k$ removed, each disk corresponding to a conical 
singularity.
By using (\ref{a.12}) we can write
\begin{equation}\label{b.20}
\int_{{\mbox{\small{reg}}} (M^2)}\, {\bf n}\, K\, dA
 = \int_{S^2} {\bf n}\, da - \sum_{k=1}^{n} \int_{D_k} {\bf n}\, da,
\end{equation} 
Because the first term in the left hand side of (\ref{b.20}) vanishes,
we get the following identity:
\begin{equation}\label{ab.21}
\int_{{\mbox{\small{reg}}} (M^2)}\, {\bf n}\, K\, dA
 + \sum_{k=1}^{n}\int_{D_k} {\bf n}\, da = 0.
\end{equation}
By taking into account (\ref{ba.2}), this identity can be also
rewritten as 
\begin{equation}\label{ab.22}
\int_{{\mbox{\small{reg}}} (M^2)}\, {\bf n}\, K\, dA
 + \sum_{k=1}^{n}\mu_k\hat{\bf n}_k = 0,
\end{equation}
were $\hat{\bf n}_k$ is the averaged normal ${\bf n}$
over the area of the disk $D_k$, see (\ref{D14}).
Now if we subtract (\ref{D3}) from (\ref{ab.22}) and use the definition
(\ref{D11b}) of the force ${\bf F}_k$ we get
\begin{equation}
\label{ab.23}
{\bf F} \equiv \sum_{k=1}^n \mu_k\, \hat{{\bf n}}_k = 
\int_{{\mbox{\small{reg}}} (M^2)}\, {\bf n}\, (1-K)\, dA\, .
\end{equation}
The constant $1$ in this equation can be any constant, because of 
(\ref{D3}), but it is here chosen to be $1$, the value of
the Gaussian curvature on the unit thorny sphere, so that the
right hand side is obviously zero for the thorny sphere.
Eq. (\ref{ab.23}) can be considered as the generalized constraint
equation for a closed thornifold whose Gaussian curvature
$K$ is not constant.

\subsection{$C$-metric example}

To illustrate the action of the generalized constraint equation
we discuss how it works for $C$-metrics.
The $C$-metric is the following solution of the Einstein equations: 
\begin{equation}\label{a.20}
ds^2=-H\, du^2 -2\, du\, dr -2\, a\, r^2\, du\, dx +r^2\, d\omega^2\, ,
\end{equation}
\begin{equation}\label{a.21}
d\omega^2 ={dx^2\over G(x)}+G(x)\, d\tilde{\phi}^2\, ,
\end{equation}
\begin{equation}\label{a.22}
H=-a^2\, r^2\, G(x-1/(ar))\, ,\hspace{0.5cm}
G(x)=1-x^2-\alpha \, x^3\, .
\end{equation}
This metric describes the gravitational field of a uniformly accelerated
black hole (see e.g. \cite{KiWa:70}).
The parameter $a$ is the acceleration, and $\alpha=r_g\, a$,
where $r_g=2M$ is the Schwarzschild gravitational radius. 

We focus our attention on the geometry of a 2-dimensional surface
$u=$const, $r$=const. Its geometry is $r^2 d\omega^2$.
It is easy to show that for $\alpha<2/3\sqrt{3}$
the equation $G(x)=0$ has 3 real roots. Two of them are negative,
say $x_3<x_2<0$, and one is positive $x_1>0$.
From now on we assume that  $x\in (x_2,x_1)$,
so that the function $G$ is positive.
One also has $G'(x_2)>0$ and $G'(x_1)<0$.
For an arbitrary period of $\tilde{\phi}$ the surface $\Sigma$
with the metric $d\omega^2$ is a thornifold
with two conical singularities at $x_1$ and $x_2$.
The singularity at $x_1$ vanishes when
\begin{equation}\label{a.23}
\tilde{\phi} =b\, \phi\equiv {2\over |G'(x_1)|}\, \phi\, ,
\hspace{0.5cm}\phi \in (0,2\pi)\, .
\end{equation}
We fix this periodicity and write the metric (\ref{a.21}) as
\begin{equation}\label{a.24}
d\omega^2 = h_{ab} dx^a dx^b = {dx^2\over G(x)}+b^2\, G(x)\, d\phi^2\, .
\end{equation}

\begin{figure}
\centerline{\epsfig{file=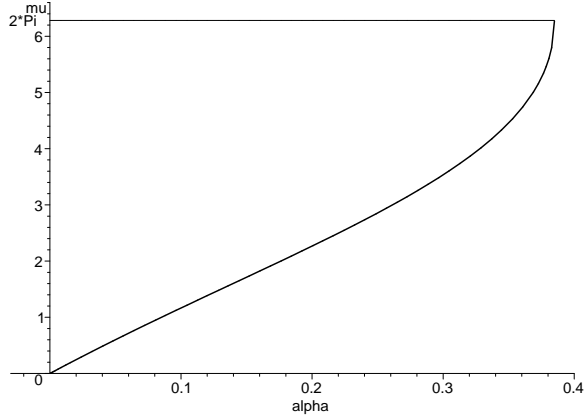, height=6.0cm}}
\caption[f1]{The angle deficit $\mu$ as a function of $\alpha$.}
\label{mu}
\end{figure}

The Gaussian curvature of $\Sigma$ is
\begin{equation}\label{a.25}
K = {1\over 2} R = -  {1\over 2} G'' = 1 + 3\, \alpha\, x\, .
\end{equation}
Integrating $K$ over the regular part of $\Sigma$, we get
\begin{equation}\label{a.26}
\int \, K\, \sqrt{h}\, d^2 x = {2\pi \over |G'(x_1)|}\left[ 
G'(x_2)+|G'(x_1)|
\right] \, .
\end{equation}
In order to satisfy the Gauss-Bonnet equation,
the conical singularity at $x_1$ must have the angle deficit
\begin{equation}\label{a.27}
\mu=2\pi \left(1-{G'(x_2)\over |G'(x_1)|}  \right)\, . 
\end{equation}
The dependence of $\mu$ on $\alpha$ can be presented in the following parametric form
\begin{equation}\label{a.27aa}
\mu= {2\pi\epsilon\, (4-5\epsilon +2\epsilon^2)\over (1-\epsilon)\, (2+\epsilon -2\epsilon^2)}\, ,\hspace{0.5cm}
\alpha={\epsilon\, (1-\epsilon)\over (1-\epsilon+\epsilon^2)^{3/2}}\, .
\end{equation}
The angle deficit $\mu$ as a function of $\alpha$ is shown
in Fig.~\ref{mu}.

The surface $\Sigma$ can be embedded into a 3-dimensional flat space
as a surface of rotation. In cylindrical coordinates the equation
of this surface is
\begin{equation}\label{a.28}
\rho=b\, \sqrt{G(x)}\,,\hspace{0.5cm}
z=-\int_{x_2}^x\, dx\, \sqrt{ {1-b^2\,G'(x)^2}/4\over G(x)} \,.
\end{equation}
Since $G'(0)=0$, one has $\rho'(0)=0$ and hence a normal vector to
$\Sigma$ at the line $x=0$ is orthogonal to the $z$-axis.
Using (\ref{a.25}) we find that above this line $K < 1$
and below it $K > 1$.

\begin{figure}
\centerline{\epsfig{file=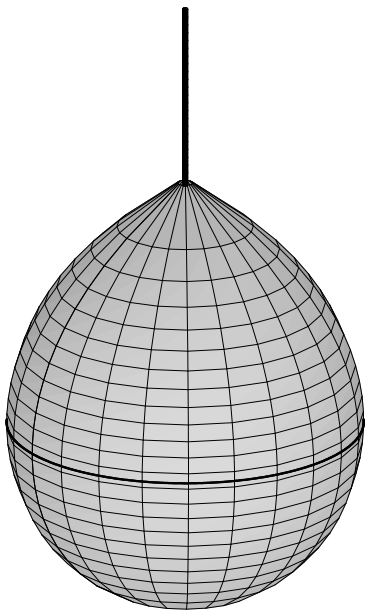, height=8.0cm}}
\caption[f1]{Embedding of $\Sigma$  into a 3 dimensional flat space ($\alpha=0.1$).
A `string' responsible for an angle deficit enters the north pole. 
A solid line on $\Sigma$ separates it into two parts, the upper one
where the Gaussian curvature is less than 1 and the lower one where
it is greater than 1.}
\label{form_C}
\end{figure} 

When the string tension is small one can apply
the generalized constraint equation to the case shown
in Fig.~\ref{form_C}.
This equation shows that as a result of the action of the
`inertial' force
of acceleration, the form of the surface is changed so that
extra positive curvature is located in the lower part of the surface,
while the extra negative  curvature is located in the upper part.
As a result, the generalized constraint equation
(\ref{ab.23}) is obeyed.

Another surface of interest is the event horizon. It is defined by the equation $H=0$, or $r=1/(a(x-x_3))$. A surface of rotation in a 3-dimensional space which has the same internal geometry as the horizon is given by equations
\begin{equation}\label{a.28a}
\rho={b\, \sqrt{G(x)}\over x-x_3}\,,\hspace{0.5cm}
z=-\int_{x_2}^x\, {dx\over x-x_3}\, \sqrt{ {1-b^2\,G'(x)^2}/4\over G(x)} \,.
\end{equation}
The form of this surface is very similar to the one shown in Fig.~\ref{form_C}.

\section{Simple examples of embedding of a thorny sphere into a
Euclidean space}

\setcounter{equation}0

When all the angle deficits are positive 
(which is the case of the most physical interest), a thorny sphere can
be  considered as a special limit of a 2-dimensional compact surface 
diffeomorphic to $S^2$ which has positive curvature and which can be embedded into a three-dimensional flat space. The curvature  is
everywhere constant except for $N_v$ localized regions $v_k$ where it 
is high. An angle deficit $\mu_k$ arises in the limit when the size of 
the region $v_k$ tends to zero while the curvature inside it grows 
infinitely, so that integral of the Gaussian curvature $K$ (one-half
the Ricci scalar curvature $R$) over this region remains finite and has
the limit $\mu_k$.

\begin{figure}
\begin{center}
\[
\begin{array}{cc}
\hspace{-1cm}
\epsfig{file=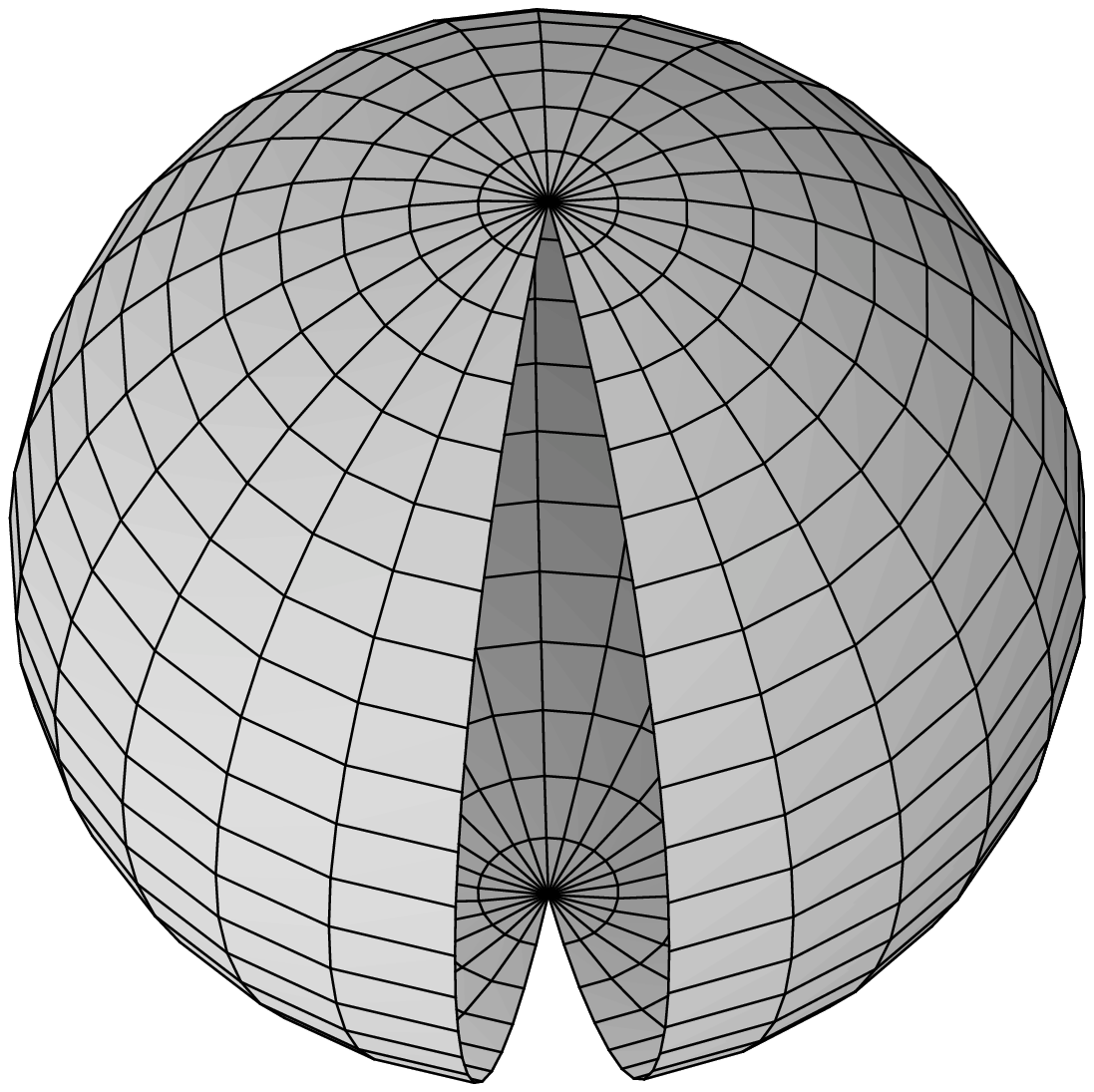, height=5.0cm}
&
\hspace{2cm}\epsfig{file=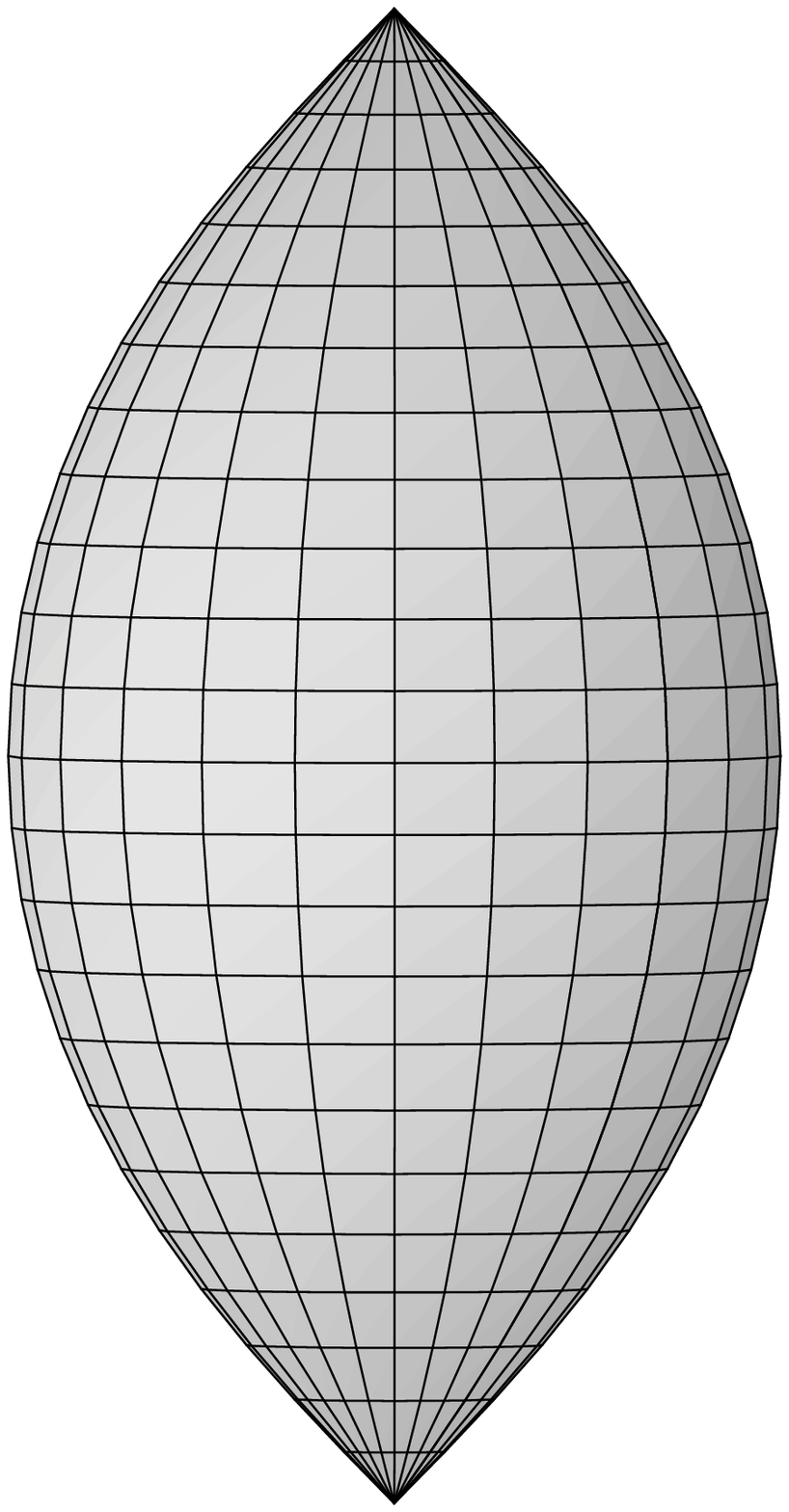, height=5.0cm}\\
\hspace{-1cm}{\bf a} & \hspace{2cm}{\bf b}
\end{array}
\]
\end{center}
\caption[f1aa]{A unit sphere with a cut (a),
and embedding of a thorny sphere with two positive angle
deficits into a 3-dimensional flat space (b).
}
\label{fB1}
\end{figure}

\begin{figure}
\centerline{\epsfig{file=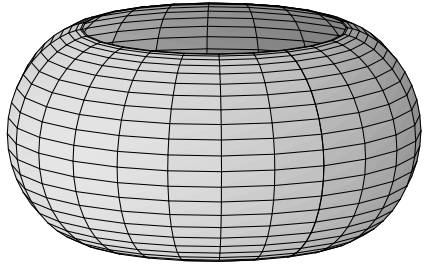, width=5.5cm}}
\vspace{1cm}
\caption[f3]{Embedding of a thorny sphere with two negative angle
deficits into a 3-dimensional flat space.}
\label{BB2}
\end{figure}

The simplest example of a thorny sphere is a sphere with 
two conical singularities located at its poles. This sphere can be 
obtained by cutting a unit sphere by planes $P_1$ and $P_2$ at angles 
$\phi=0$ and $\phi=\mu$ and gluing the cuts together
(see Fig.~\ref{fB1} (a)). 
It can also be obtained as the geometry on the surface of rotation 
embedded in a flat 3 space (see e.g. \cite{Bl}).
This surface is obtained by a rotation 
around the $z$-axis of the following 
meridianal curve 
\begin{equation}\label{1a}
x=a\cos \sigma\, ,\hspace{0.5cm}
y=0\, ,\hspace{0.5cm}
z=\int_0^{\sigma}\sqrt{1-a^2\sin^2\sigma}\, d\sigma\, ,
\hspace{0.5cm}|\sigma|\le \pi/2\, .
\end{equation}
Here $a\le 1$.  For $a=1$ the angle deficit vanishes. 
In the general case the angle deficit is $2\pi(1-a)$. This surface is 
shown in Fig.~\ref{fB1} (b).
A similar surface of rotation of constant curvature 
for a negative angle deficit is shown at Fig.~\ref{BB2}.
(Note that it is impossible to embed the entirety
of this surface in 3-dimensional flat space in an
axially symmetric way, so the embedding
stops before one gets to the conical singularities with negative
deficit angles.)
The corresponding
equations of the meridian curve which generate this rotational surface
are
\begin{equation}\label{1b}
x=\sqrt{a^2-\sin^2{\sigma}}\, ,\hspace{0.5cm}
y=0\, ,\hspace{0.5cm}
z=\int_0^{\sigma}{\cos^2 \sigma\over\sqrt{a^2-\sin^2\sigma}}\,
d\sigma\, .
\end{equation}

\end{document}